\documentclass[preprint,aps,showpacs,eqsecnum,superscriptaddress]{revtex4}
\usepackage{graphicx,amsfonts}
\usepackage{dcolumn}
\usepackage{bm}

\def\der{{\buildrel\,\leftarrow\over D}\!\!}

\def\a{\alpha}
\def\b{\beta}
\def\d{\delta}

\def\f{\phi}

\def\k{\kappa}
\def\l{\lambda}
\def\m{\mu}
\def\n{\nu}
\def\o{\omega}

\def\q{\theta}
\def\r{\rho}
\def\s{\sigma}
\def\t{\tau}

\def\x{\xi}

\def\L{\Lambda}

\def\S{\Sigma}

\newcommand{\hf}{\frac12}
\newcommand{\be}{\begin{equation}}
\newcommand{\ee}{\end{equation}}
\newcommand{\bea}{\begin{eqnarray}}
\newcommand{\eea}{\end{eqnarray}}

\def \N {{\cal N}}

\def\d{\partial}

\def \N {{\cal N }}

\begin{document}
\immediate\write16{<<WARNING: LINEDRAW macros work with emTeX-dvivers
                    and other drivers supporting emTeX \special's
                    (dviscr, dvihplj, dvidot, dvips, dviwin, etc.) >>}
\newdimen\Lengthunit       \Lengthunit  = 1.5cm
\newcount\Nhalfperiods     \Nhalfperiods= 9
\newcount\magnitude        \magnitude = 1000

\catcode`\*=11
\newdimen\L*   \newdimen\d*   \newdimen\d**
\newdimen\dm*  \newdimen\dd*  \newdimen\dt*
\newdimen\a*   \newdimen\b*   \newdimen\c*
\newdimen\a**  \newdimen\b**
\newdimen\xL*  \newdimen\yL*
\newdimen\rx*  \newdimen\ry*
\newdimen\tmp* \newdimen\linwid*

\newcount\k*   \newcount\l*   \newcount\m*
\newcount\k**  \newcount\l**  \newcount\m**
\newcount\n*   \newcount\dn*  \newcount\r*
\newcount\N*   \newcount\*one \newcount\*two  \*one=1 \*two=2
\newcount\*ths \*ths=1000
\newcount\angle*  \newcount\q*  \newcount\q**
\newcount\angle** \angle**=0
\newcount\sc*     \sc*=0

\newtoks\cos*  \cos*={1}
\newtoks\sin*  \sin*={0}

\catcode`\[=13

\def\rotate(#1){\advance\angle**#1\angle*=\angle**
\q**=\angle*\ifnum\q**<0\q**=-\q**\fi
\ifnum\q**>360\q*=\angle*\divide\q*360\multiply\q*360\advance\angle*-\q*\fi
\ifnum\angle*<0\advance\angle*360\fi\q**=\angle*\divide\q**90\q**=\q**
\def\sgcos*{+}\def\sgsin*{+}\relax
\ifcase\q**\or
 \def\sgcos*{-}\def\sgsin*{+}\or
 \def\sgcos*{-}\def\sgsin*{-}\or
 \def\sgcos*{+}\def\sgsin*{-}\else\fi
\q*=\q**
\multiply\q*90\advance\angle*-\q*
\ifnum\angle*>45\sc*=1\angle*=-\angle*\advance\angle*90\else\sc*=0\fi
\def[##1,##2]{\ifnum\sc*=0\relax
\edef\cs*{\sgcos*.##1}\edef\sn*{\sgsin*.##2}\ifcase\q**\or
 \edef\cs*{\sgcos*.##2}\edef\sn*{\sgsin*.##1}\or
 \edef\cs*{\sgcos*.##1}\edef\sn*{\sgsin*.##2}\or
 \edef\cs*{\sgcos*.##2}\edef\sn*{\sgsin*.##1}\else\fi\else
\edef\cs*{\sgcos*.##2}\edef\sn*{\sgsin*.##1}\ifcase\q**\or
 \edef\cs*{\sgcos*.##1}\edef\sn*{\sgsin*.##2}\or
 \edef\cs*{\sgcos*.##2}\edef\sn*{\sgsin*.##1}\or
 \edef\cs*{\sgcos*.##1}\edef\sn*{\sgsin*.##2}\else\fi\fi
\cos*={\cs*}\sin*={\sn*}\global\edef\gcos*{\cs*}\global\edef\gsin*{\sn*}}\relax
\ifcase\angle*[9999,0]\or
[999,017]\or[999,034]\or[998,052]\or[997,069]\or[996,087]\or
[994,104]\or[992,121]\or[990,139]\or[987,156]\or[984,173]\or
[981,190]\or[978,207]\or[974,224]\or[970,241]\or[965,258]\or
[961,275]\or[956,292]\or[951,309]\or[945,325]\or[939,342]\or
[933,358]\or[927,374]\or[920,390]\or[913,406]\or[906,422]\or
[898,438]\or[891,453]\or[882,469]\or[874,484]\or[866,499]\or
[857,515]\or[848,529]\or[838,544]\or[829,559]\or[819,573]\or
[809,587]\or[798,601]\or[788,615]\or[777,629]\or[766,642]\or
[754,656]\or[743,669]\or[731,681]\or[719,694]\or[707,707]\or
\else[9999,0]\fi}

\catcode`\[=12

\def\GRAPH(hsize=#1)#2{\hbox to #1\Lengthunit{#2\hss}}

\def\Linewidth#1{\global\linwid*=#1\relax
\global\divide\linwid*10\global\multiply\linwid*\mag
\global\divide\linwid*100\special{em:linewidth \the\linwid*}}

\Linewidth{.4pt}
\def\sm*{\special{em:moveto}}
\def\sl*{\special{em:lineto}}
\let\moveto=\sm*
\let\lineto=\sl*
\newbox\spm*   \newbox\spl*
\setbox\spm*\hbox{\sm*}
\setbox\spl*\hbox{\sl*}

\def\mov#1(#2,#3)#4{\rlap{\L*=#1\Lengthunit
\xL*=#2\L* \yL*=#3\L*
\xL*=\xscale\xL* \yL*=\yscale\yL*
\rx* \the\cos*\xL* \tmp* \the\sin*\yL* \advance\rx*-\tmp*
\ry* \the\cos*\yL* \tmp* \the\sin*\xL* \advance\ry*\tmp*
\kern\rx*\raise\ry*\hbox{#4}}}

\def\rmov*(#1,#2)#3{\rlap{\xL*=#1\yL*=#2\relax
\rx* \the\cos*\xL* \tmp* \the\sin*\yL* \advance\rx*-\tmp*
\ry* \the\cos*\yL* \tmp* \the\sin*\xL* \advance\ry*\tmp*
\kern\rx*\raise\ry*\hbox{#3}}}

\def\lin#1(#2,#3){\rlap{\sm*\mov#1(#2,#3){\sl*}}}

\def\arr*(#1,#2,#3){\rmov*(#1\dd*,#1\dt*){\sm*
\rmov*(#2\dd*,#2\dt*){\rmov*(#3\dt*,-#3\dd*){\sl*}}\sm*
\rmov*(#2\dd*,#2\dt*){\rmov*(-#3\dt*,#3\dd*){\sl*}}}}

\def\arrow#1(#2,#3){\rlap{\lin#1(#2,#3)\mov#1(#2,#3){\relax
\d**=-.012\Lengthunit\dd*=#2\d**\dt*=#3\d**
\arr*(1,10,4)\arr*(3,8,4)\arr*(4.8,4.2,3)}}}

\def\arrlin#1(#2,#3){\rlap{\L*=#1\Lengthunit\L*=.5\L*
\lin#1(#2,#3)\rmov*(#2\L*,#3\L*){\arrow.1(#2,#3)}}}

\def\dasharrow#1(#2,#3){\rlap{{\Lengthunit=0.9\Lengthunit
\dashlin#1(#2,#3)\mov#1(#2,#3){\sm*}}\mov#1(#2,#3){\sl*
\d**=-.012\Lengthunit\dd*=#2\d**\dt*=#3\d**
\arr*(1,10,4)\arr*(3,8,4)\arr*(4.8,4.2,3)}}}

\def\clap#1{\hbox to 0pt{\hss #1\hss}}

\def\ind(#1,#2)#3{\rlap{\L*=.1\Lengthunit
\xL*=#1\L* \yL*=#2\L*
\rx* \the\cos*\xL* \tmp* \the\sin*\yL* \advance\rx*-\tmp*
\ry* \the\cos*\yL* \tmp* \the\sin*\xL* \advance\ry*\tmp*
\kern\rx*\raise\ry*\hbox{\lower2pt\clap{$#3$}}}}

\def\sh*(#1,#2)#3{\rlap{\dm*=\the\n*\d**
\xL*=\xscale\dm* \yL*=\yscale\dm* \xL*=#1\xL* \yL*=#2\yL*
\rx* \the\cos*\xL* \tmp* \the\sin*\yL* \advance\rx*-\tmp*
\ry* \the\cos*\yL* \tmp* \the\sin*\xL* \advance\ry*\tmp*
\kern\rx*\raise\ry*\hbox{#3}}}

\def\calcnum*#1(#2,#3){\a*=1000sp\b*=1000sp\a*=#2\a*\b*=#3\b*
\ifdim\a*<0pt\a*-\a*\fi\ifdim\b*<0pt\b*-\b*\fi
\ifdim\a*>\b*\c*=.96\a*\advance\c*.4\b*
\else\c*=.96\b*\advance\c*.4\a*\fi
\k*\a*\multiply\k*\k*\l*\b*\multiply\l*\l*
\m*\k*\advance\m*\l*\n*\c*\r*\n*\multiply\n*\n*
\dn*\m*\advance\dn*-\n*\divide\dn*2\divide\dn*\r*
\advance\r*\dn*
\c*=\the\Nhalfperiods5sp\c*=#1\c*\ifdim\c*<0pt\c*-\c*\fi
\multiply\c*\r*\N*\c*\divide\N*10000}

\def\dashlin#1(#2,#3){\rlap{\calcnum*#1(#2,#3)\relax
\d**=#1\Lengthunit\ifdim\d**<0pt\d**-\d**\fi
\divide\N*2\multiply\N*2\advance\N*\*one
\divide\d**\N*\sm*\n*\*one\sh*(#2,#3){\sl*}\loop
\advance\n*\*one\sh*(#2,#3){\sm*}\advance\n*\*one
\sh*(#2,#3){\sl*}\ifnum\n*<\N*\repeat}}

\def\dashdotlin#1(#2,#3){\rlap{\calcnum*#1(#2,#3)\relax
\d**=#1\Lengthunit\ifdim\d**<0pt\d**-\d**\fi
\divide\N*2\multiply\N*2\advance\N*1\multiply\N*2\relax
\divide\d**\N*\sm*\n*\*two\sh*(#2,#3){\sl*}\loop
\advance\n*\*one\sh*(#2,#3){\kern-1.48pt\lower.5pt\hbox{\rm.}}\relax
\advance\n*\*one\sh*(#2,#3){\sm*}\advance\n*\*two
\sh*(#2,#3){\sl*}\ifnum\n*<\N*\repeat}}

\def\shl*(#1,#2)#3{\kern#1#3\lower#2#3\hbox{\unhcopy\spl*}}

\def\trianglin#1(#2,#3){\rlap{\toks0={#2}\toks1={#3}\calcnum*#1(#2,#3)\relax
\dd*=.57\Lengthunit\dd*=#1\dd*\divide\dd*\N*
\divide\dd*\*ths \multiply\dd*\magnitude
\d**=#1\Lengthunit\ifdim\d**<0pt\d**-\d**\fi
\multiply\N*2\divide\d**\N*\sm*\n*\*one\loop
\shl**{\dd*}\dd*-\dd*\advance\n*2\relax
\ifnum\n*<\N*\repeat\n*\N*\shl**{0pt}}}

\def\wavelin#1(#2,#3){\rlap{\toks0={#2}\toks1={#3}\calcnum*#1(#2,#3)\relax
\dd*=.23\Lengthunit\dd*=#1\dd*\divide\dd*\N*
\divide\dd*\*ths \multiply\dd*\magnitude
\d**=#1\Lengthunit\ifdim\d**<0pt\d**-\d**\fi
\multiply\N*4\divide\d**\N*\sm*\n*\*one\loop
\shl**{\dd*}\dt*=1.3\dd*\advance\n*\*one
\shl**{\dt*}\advance\n*\*one
\shl**{\dd*}\advance\n*\*two
\dd*-\dd*\ifnum\n*<\N*\repeat\n*\N*\shl**{0pt}}}

\def\w*lin(#1,#2){\rlap{\toks0={#1}\toks1={#2}\d**=\Lengthunit\dd*=-.12\d**
\divide\dd*\*ths \multiply\dd*\magnitude
\N*8\divide\d**\N*\sm*\n*\*one\loop
\shl**{\dd*}\dt*=1.3\dd*\advance\n*\*one
\shl**{\dt*}\advance\n*\*one
\shl**{\dd*}\advance\n*\*one
\shl**{0pt}\dd*-\dd*\advance\n*1\ifnum\n*<\N*\repeat}}

\def\l*arc(#1,#2)[#3][#4]{\rlap{\toks0={#1}\toks1={#2}\d**=\Lengthunit
\dd*=#3.037\d**\dd*=#4\dd*\dt*=#3.049\d**\dt*=#4\dt*\ifdim\d**>10mm\relax
\d**=.25\d**\n*\*one\shl**{-\dd*}\n*\*two\shl**{-\dt*}\n*3\relax
\shl**{-\dd*}\n*4\relax\shl**{0pt}\else
\ifdim\d**>5mm\d**=.5\d**\n*\*one\shl**{-\dt*}\n*\*two
\shl**{0pt}\else\n*\*one\shl**{0pt}\fi\fi}}

\def\d*arc(#1,#2)[#3][#4]{\rlap{\toks0={#1}\toks1={#2}\d**=\Lengthunit
\dd*=#3.037\d**\dd*=#4\dd*\d**=.25\d**\sm*\n*\*one\shl**{-\dd*}\relax
\n*3\relax\sh*(#1,#2){\xL*=\xscale\dd*\yL*=\yscale\dd*
\kern#2\xL*\lower#1\yL*\hbox{\sm*}}\n*4\relax\shl**{0pt}}}

\def\shl**#1{\c*=\the\n*\d**\d*=#1\relax
\a*=\the\toks0\c*\b*=\the\toks1\d*\advance\a*-\b*
\b*=\the\toks1\c*\d*=\the\toks0\d*\advance\b*\d*
\a*=\xscale\a*\b*=\yscale\b*
\rx* \the\cos*\a* \tmp* \the\sin*\b* \advance\rx*-\tmp*
\ry* \the\cos*\b* \tmp* \the\sin*\a* \advance\ry*\tmp*
\raise\ry*\rlap{\kern\rx*\unhcopy\spl*}}

\def\wlin*#1(#2,#3)[#4]{\rlap{\toks0={#2}\toks1={#3}\relax
\c*=#1\l*\c*\c*=.01\Lengthunit\m*\c*\divide\l*\m*
\c*=\the\Nhalfperiods5sp\multiply\c*\l*\N*\c*\divide\N*\*ths
\divide\N*2\multiply\N*2\advance\N*\*one
\dd*=.002\Lengthunit\dd*=#4\dd*\multiply\dd*\l*\divide\dd*\N*
\divide\dd*\*ths \multiply\dd*\magnitude
\d**=#1\multiply\N*4\divide\d**\N*\sm*\n*\*one\loop
\shl**{\dd*}\dt*=1.3\dd*\advance\n*\*one
\shl**{\dt*}\advance\n*\*one
\shl**{\dd*}\advance\n*\*two
\dd*-\dd*\ifnum\n*<\N*\repeat\n*\N*\shl**{0pt}}}

\def\wavebox#1{\setbox0\hbox{#1}\relax
\a*=\wd0\advance\a*14pt\b*=\ht0\advance\b*\dp0\advance\b*14pt\relax
\hbox{\kern9pt\relax
\rmov*(0pt,\ht0){\rmov*(-7pt,7pt){\wlin*\a*(1,0)[+]\wlin*\b*(0,-1)[-]}}\relax
\rmov*(\wd0,-\dp0){\rmov*(7pt,-7pt){\wlin*\a*(-1,0)[+]\wlin*\b*(0,1)[-]}}\relax
\box0\kern9pt}}

\def\rectangle#1(#2,#3){\relax
\lin#1(#2,0)\lin#1(0,#3)\mov#1(0,#3){\lin#1(#2,0)}\mov#1(#2,0){\lin#1(0,#3)}}

\def\dashrectangle#1(#2,#3){\dashlin#1(#2,0)\dashlin#1(0,#3)\relax
\mov#1(0,#3){\dashlin#1(#2,0)}\mov#1(#2,0){\dashlin#1(0,#3)}}

\def\waverectangle#1(#2,#3){\L*=#1\Lengthunit\a*=#2\L*\b*=#3\L*
\ifdim\a*<0pt\a*-\a*\def\x*{-1}\else\def\x*{1}\fi
\ifdim\b*<0pt\b*-\b*\def\y*{-1}\else\def\y*{1}\fi
\wlin*\a*(\x*,0)[-]\wlin*\b*(0,\y*)[+]\relax
\mov#1(0,#3){\wlin*\a*(\x*,0)[+]}\mov#1(#2,0){\wlin*\b*(0,\y*)[-]}}

\def\calcparab*{\ifnum\n*>\m*\k*\N*\advance\k*-\n*\else\k*\n*\fi
\a*=\the\k* sp\a*=10\a*\b*\dm*\advance\b*-\a*\k*\b*
\a*=\the\*ths\b*\divide\a*\l*\multiply\a*\k*
\divide\a*\l*\k*\*ths\r*\a*\advance\k*-\r*\dt*=\the\k*\L*}

\def\arcto#1(#2,#3)[#4]{\rlap{\toks0={#2}\toks1={#3}\calcnum*#1(#2,#3)\relax
\dm*=135sp\dm*=#1\dm*\d**=#1\Lengthunit\ifdim\dm*<0pt\dm*-\dm*\fi
\multiply\dm*\r*\a*=.3\dm*\a*=#4\a*\ifdim\a*<0pt\a*-\a*\fi
\advance\dm*\a*\N*\dm*\divide\N*10000\relax
\divide\N*2\multiply\N*2\advance\N*\*one
\L*=-.25\d**\L*=#4\L*\divide\d**\N*\divide\L*\*ths
\m*\N*\divide\m*2\dm*=\the\m*5sp\l*\dm*\sm*\n*\*one\loop
\calcparab*\shl**{-\dt*}\advance\n*1\ifnum\n*<\N*\repeat}}

\def\arrarcto#1(#2,#3)[#4]{\L*=#1\Lengthunit\L*=.54\L*
\arcto#1(#2,#3)[#4]\rmov*(#2\L*,#3\L*){\d*=.457\L*\d*=#4\d*\d**-\d*
\rmov*(#3\d**,#2\d*){\arrow.02(#2,#3)}}}

\def\dasharcto#1(#2,#3)[#4]{\rlap{\toks0={#2}\toks1={#3}\relax
\calcnum*#1(#2,#3)\dm*=\the\N*5sp\a*=.3\dm*\a*=#4\a*\ifdim\a*<0pt\a*-\a*\fi
\advance\dm*\a*\N*\dm*
\divide\N*20\multiply\N*2\advance\N*1\d**=#1\Lengthunit
\L*=-.25\d**\L*=#4\L*\divide\d**\N*\divide\L*\*ths
\m*\N*\divide\m*2\dm*=\the\m*5sp\l*\dm*
\sm*\n*\*one\loop\calcparab*
\shl**{-\dt*}\advance\n*1\ifnum\n*>\N*\else\calcparab*
\sh*(#2,#3){\xL*=#3\dt* \yL*=#2\dt*
\rx* \the\cos*\xL* \tmp* \the\sin*\yL* \advance\rx*\tmp*
\ry* \the\cos*\yL* \tmp* \the\sin*\xL* \advance\ry*-\tmp*
\kern\rx*\lower\ry*\hbox{\sm*}}\fi
\advance\n*1\ifnum\n*<\N*\repeat}}

\def\*shl*#1{\c*=\the\n*\d**\advance\c*#1\a**\d*\dt*\advance\d*#1\b**
\a*=\the\toks0\c*\b*=\the\toks1\d*\advance\a*-\b*
\b*=\the\toks1\c*\d*=\the\toks0\d*\advance\b*\d*
\rx* \the\cos*\a* \tmp* \the\sin*\b* \advance\rx*-\tmp*
\ry* \the\cos*\b* \tmp* \the\sin*\a* \advance\ry*\tmp*
\raise\ry*\rlap{\kern\rx*\unhcopy\spl*}}

\def\calcnormal*#1{\b**=10000sp\a**\b**\k*\n*\advance\k*-\m*
\multiply\a**\k*\divide\a**\m*\a**=#1\a**\ifdim\a**<0pt\a**-\a**\fi
\ifdim\a**>\b**\d*=.96\a**\advance\d*.4\b**
\else\d*=.96\b**\advance\d*.4\a**\fi
\d*=.01\d*\r*\d*\divide\a**\r*\divide\b**\r*
\ifnum\k*<0\a**-\a**\fi\d*=#1\d*\ifdim\d*<0pt\b**-\b**\fi
\k*\a**\a**=\the\k*\dd*\k*\b**\b**=\the\k*\dd*}

\def\wavearcto#1(#2,#3)[#4]{\rlap{\toks0={#2}\toks1={#3}\relax
\calcnum*#1(#2,#3)\c*=\the\N*5sp\a*=.4\c*\a*=#4\a*\ifdim\a*<0pt\a*-\a*\fi
\advance\c*\a*\N*\c*\divide\N*20\multiply\N*2\advance\N*-1\multiply\N*4\relax
\d**=#1\Lengthunit\dd*=.012\d**
\divide\dd*\*ths \multiply\dd*\magnitude
\ifdim\d**<0pt\d**-\d**\fi\L*=.25\d**
\divide\d**\N*\divide\dd*\N*\L*=#4\L*\divide\L*\*ths
\m*\N*\divide\m*2\dm*=\the\m*0sp\l*\dm*
\sm*\n*\*one\loop\calcnormal*{#4}\calcparab*
\*shl*{1}\advance\n*\*one\calcparab*
\*shl*{1.3}\advance\n*\*one\calcparab*
\*shl*{1}\advance\n*2\dd*-\dd*\ifnum\n*<\N*\repeat\n*\N*\shl**{0pt}}}

\def\triangarcto#1(#2,#3)[#4]{\rlap{\toks0={#2}\toks1={#3}\relax
\calcnum*#1(#2,#3)\c*=\the\N*5sp\a*=.4\c*\a*=#4\a*\ifdim\a*<0pt\a*-\a*\fi
\advance\c*\a*\N*\c*\divide\N*20\multiply\N*2\advance\N*-1\multiply\N*2\relax
\d**=#1\Lengthunit\dd*=.012\d**
\divide\dd*\*ths \multiply\dd*\magnitude
\ifdim\d**<0pt\d**-\d**\fi\L*=.25\d**
\divide\d**\N*\divide\dd*\N*\L*=#4\L*\divide\L*\*ths
\m*\N*\divide\m*2\dm*=\the\m*0sp\l*\dm*
\sm*\n*\*one\loop\calcnormal*{#4}\calcparab*
\*shl*{1}\advance\n*2\dd*-\dd*\ifnum\n*<\N*\repeat\n*\N*\shl**{0pt}}}

\def\hr*#1{\L*=\xscale\Lengthunit\ifnum
\angle**=0\clap{\vrule width#1\L* height.1pt}\else
\L*=#1\L*\L*=.5\L*\rmov*(-\L*,0pt){\sm*}\rmov*(\L*,0pt){\sl*}\fi}

\def\shade#1[#2]{\rlap{\Lengthunit=#1\Lengthunit
\special{em:linewidth .001pt}\relax
\mov(0,#2.05){\hr*{.994}}\mov(0,#2.1){\hr*{.980}}\relax
\mov(0,#2.15){\hr*{.953}}\mov(0,#2.2){\hr*{.916}}\relax
\mov(0,#2.25){\hr*{.867}}\mov(0,#2.3){\hr*{.798}}\relax
\mov(0,#2.35){\hr*{.715}}\mov(0,#2.4){\hr*{.603}}\relax
\mov(0,#2.45){\hr*{.435}}\special{em:linewidth \the\linwid*}}}

\def\dshade#1[#2]{\rlap{\special{em:linewidth .001pt}\relax
\Lengthunit=#1\Lengthunit\if#2-\def\t*{+}\else\def\t*{-}\fi
\mov(0,\t*.025){\relax
\mov(0,#2.05){\hr*{.995}}\mov(0,#2.1){\hr*{.988}}\relax
\mov(0,#2.15){\hr*{.969}}\mov(0,#2.2){\hr*{.937}}\relax
\mov(0,#2.25){\hr*{.893}}\mov(0,#2.3){\hr*{.836}}\relax
\mov(0,#2.35){\hr*{.760}}\mov(0,#2.4){\hr*{.662}}\relax
\mov(0,#2.45){\hr*{.531}}\mov(0,#2.5){\hr*{.320}}\relax
\special{em:linewidth \the\linwid*}}}}

\def\vdot{\rlap{\kern-1.9pt\lower1.8pt\hbox{$\scriptstyle\bullet$}}}
\def\vtimes{\rlap{\kern-3pt\lower1.8pt\hbox{$\scriptstyle\times$}}}
\def\vDot{\rlap{\kern-2.3pt\lower2.7pt\hbox{$\bullet$}}}
\def\vTimes{\rlap{\kern-3.6pt\lower2.4pt\hbox{$\times$}}}

\def\arc(#1)[#2,#3]{{\k*=#2\l*=#3\m*=\l*
\advance\m*-6\ifnum\k*>\l*\relax\else
{\rotate(#2)\mov(#1,0){\sm*}}\loop
\ifnum\k*<\m*\advance\k*5{\rotate(\k*)\mov(#1,0){\sl*}}\repeat
{\rotate(#3)\mov(#1,0){\sl*}}\fi}}

\def\dasharc(#1)[#2,#3]{{\k**=#2\n*=#3\advance\n*-1\advance\n*-\k**
\L*=1000sp\L*#1\L* \multiply\L*\n* \multiply\L*\Nhalfperiods
\divide\L*57\N*\L* \divide\N*2000\ifnum\N*=0\N*1\fi
\r*\n*  \divide\r*\N* \ifnum\r*<2\r*2\fi
\m**\r* \divide\m**2 \l**\r* \advance\l**-\m** \N*\n* \divide\N*\r*
\k**\r* \multiply\k**\N* \dn*\n* \advance\dn*-\k** \divide\dn*2\advance\dn*\*one
\r*\l** \divide\r*2\advance\dn*\r* \advance\N*-2\k**#2\relax
\ifnum\l**<6{\rotate(#2)\mov(#1,0){\sm*}}\advance\k**\dn*
{\rotate(\k**)\mov(#1,0){\sl*}}\advance\k**\m**
{\rotate(\k**)\mov(#1,0){\sm*}}\loop
\advance\k**\l**{\rotate(\k**)\mov(#1,0){\sl*}}\advance\k**\m**
{\rotate(\k**)\mov(#1,0){\sm*}}\advance\N*-1\ifnum\N*>0\repeat
{\rotate(#3)\mov(#1,0){\sl*}}\else\advance\k**\dn*
\arc(#1)[#2,\k**]\loop\advance\k**\m** \r*\k**
\advance\k**\l** {\arc(#1)[\r*,\k**]}\relax
\advance\N*-1\ifnum\N*>0\repeat
\advance\k**\m**\arc(#1)[\k**,#3]\fi}}

\def\triangarc#1(#2)[#3,#4]{{\k**=#3\n*=#4\advance\n*-\k**
\L*=1000sp\L*#2\L* \multiply\L*\n* \multiply\L*\Nhalfperiods
\divide\L*57\N*\L* \divide\N*1000\ifnum\N*=0\N*1\fi
\d**=#2\Lengthunit \d*\d** \divide\d*57\multiply\d*\n*
\r*\n*  \divide\r*\N* \ifnum\r*<2\r*2\fi
\m**\r* \divide\m**2 \l**\r* \advance\l**-\m** \N*\n* \divide\N*\r*
\dt*\d* \divide\dt*\N* \dt*.5\dt* \dt*#1\dt*
\divide\dt*1000\multiply\dt*\magnitude
\k**\r* \multiply\k**\N* \dn*\n* \advance\dn*-\k** \divide\dn*2\relax
\r*\l** \divide\r*2\advance\dn*\r* \advance\N*-1\k**#3\relax
{\rotate(#3)\mov(#2,0){\sm*}}\advance\k**\dn*
{\rotate(\k**)\mov(#2,0){\sl*}}\advance\k**-\m**\advance\l**\m**\loop\dt*-\dt*
\d*\d** \advance\d*\dt*
\advance\k**\l**{\rotate(\k**)\rmov*(\d*,0pt){\sl*}}%
\advance\N*-1\ifnum\N*>0\repeat\advance\k**\m**
{\rotate(\k**)\mov(#2,0){\sl*}}{\rotate(#4)\mov(#2,0){\sl*}}}}

\def\wavearc#1(#2)[#3,#4]{{\k**=#3\n*=#4\advance\n*-\k**
\L*=4000sp\L*#2\L* \multiply\L*\n* \multiply\L*\Nhalfperiods
\divide\L*57\N*\L* \divide\N*1000\ifnum\N*=0\N*1\fi
\d**=#2\Lengthunit \d*\d** \divide\d*57\multiply\d*\n*
\r*\n*  \divide\r*\N* \ifnum\r*=0\r*1\fi
\m**\r* \divide\m**2 \l**\r* \advance\l**-\m** \N*\n* \divide\N*\r*
\dt*\d* \divide\dt*\N* \dt*.7\dt* \dt*#1\dt*
\divide\dt*1000\multiply\dt*\magnitude
\k**\r* \multiply\k**\N* \dn*\n* \advance\dn*-\k** \divide\dn*2\relax
\divide\N*4\advance\N*-1\k**#3\relax
{\rotate(#3)\mov(#2,0){\sm*}}\advance\k**\dn*
{\rotate(\k**)\mov(#2,0){\sl*}}\advance\k**-\m**\advance\l**\m**\loop\dt*-\dt*
\d*\d** \advance\d*\dt* \dd*\d** \advance\dd*1.3\dt*
\advance\k**\r*{\rotate(\k**)\rmov*(\d*,0pt){\sl*}}\relax
\advance\k**\r*{\rotate(\k**)\rmov*(\dd*,0pt){\sl*}}\relax
\advance\k**\r*{\rotate(\k**)\rmov*(\d*,0pt){\sl*}}\relax
\advance\k**\r*
\advance\N*-1\ifnum\N*>0\repeat\advance\k**\m**
{\rotate(\k**)\mov(#2,0){\sl*}}{\rotate(#4)\mov(#2,0){\sl*}}}}

\def\gmov*#1(#2,#3)#4{\rlap{\L*=#1\Lengthunit
\xL*=#2\L* \yL*=#3\L*
\rx* \gcos*\xL* \tmp* \gsin*\yL* \advance\rx*-\tmp*
\ry* \gcos*\yL* \tmp* \gsin*\xL* \advance\ry*\tmp*
\rx*=\xscale\rx* \ry*=\yscale\ry*
\xL* \the\cos*\rx* \tmp* \the\sin*\ry* \advance\xL*-\tmp*
\yL* \the\cos*\ry* \tmp* \the\sin*\rx* \advance\yL*\tmp*
\kern\xL*\raise\yL*\hbox{#4}}}

\def\rgmov*(#1,#2)#3{\rlap{\xL*#1\yL*#2\relax
\rx* \gcos*\xL* \tmp* \gsin*\yL* \advance\rx*-\tmp*
\ry* \gcos*\yL* \tmp* \gsin*\xL* \advance\ry*\tmp*
\rx*=\xscale\rx* \ry*=\yscale\ry*
\xL* \the\cos*\rx* \tmp* \the\sin*\ry* \advance\xL*-\tmp*
\yL* \the\cos*\ry* \tmp* \the\sin*\rx* \advance\yL*\tmp*
\kern\xL*\raise\yL*\hbox{#3}}}

\def\Earc(#1)[#2,#3][#4,#5]{{\k*=#2\l*=#3\m*=\l*
\advance\m*-6\ifnum\k*>\l*\relax\else\def\xscale{#4}\def\yscale{#5}\relax
{\angle**0\rotate(#2)}\gmov*(#1,0){\sm*}\loop
\ifnum\k*<\m*\advance\k*5\relax
{\angle**0\rotate(\k*)}\gmov*(#1,0){\sl*}\repeat
{\angle**0\rotate(#3)}\gmov*(#1,0){\sl*}\relax
\def\xscale{1}\def\yscale{1}\fi}}

\def\dashEarc(#1)[#2,#3][#4,#5]{{\k**=#2\n*=#3\advance\n*-1\advance\n*-\k**
\L*=1000sp\L*#1\L* \multiply\L*\n* \multiply\L*\Nhalfperiods
\divide\L*57\N*\L* \divide\N*2000\ifnum\N*=0\N*1\fi
\r*\n*  \divide\r*\N* \ifnum\r*<2\r*2\fi
\m**\r* \divide\m**2 \l**\r* \advance\l**-\m** \N*\n* \divide\N*\r*
\k**\r*\multiply\k**\N* \dn*\n* \advance\dn*-\k** \divide\dn*2\advance\dn*\*one
\r*\l** \divide\r*2\advance\dn*\r* \advance\N*-2\k**#2\relax
\ifnum\l**<6\def\xscale{#4}\def\yscale{#5}\relax
{\angle**0\rotate(#2)}\gmov*(#1,0){\sm*}\advance\k**\dn*
{\angle**0\rotate(\k**)}\gmov*(#1,0){\sl*}\advance\k**\m**
{\angle**0\rotate(\k**)}\gmov*(#1,0){\sm*}\loop
\advance\k**\l**{\angle**0\rotate(\k**)}\gmov*(#1,0){\sl*}\advance\k**\m**
{\angle**0\rotate(\k**)}\gmov*(#1,0){\sm*}\advance\N*-1\ifnum\N*>0\repeat
{\angle**0\rotate(#3)}\gmov*(#1,0){\sl*}\def\xscale{1}\def\yscale{1}\else
\advance\k**\dn* \Earc(#1)[#2,\k**][#4,#5]\loop\advance\k**\m** \r*\k**
\advance\k**\l** {\Earc(#1)[\r*,\k**][#4,#5]}\relax
\advance\N*-1\ifnum\N*>0\repeat
\advance\k**\m**\Earc(#1)[\k**,#3][#4,#5]\fi}}

\def\triangEarc#1(#2)[#3,#4][#5,#6]{{\k**=#3\n*=#4\advance\n*-\k**
\L*=1000sp\L*#2\L* \multiply\L*\n* \multiply\L*\Nhalfperiods
\divide\L*57\N*\L* \divide\N*1000\ifnum\N*=0\N*1\fi
\d**=#2\Lengthunit \d*\d** \divide\d*57\multiply\d*\n*
\r*\n*  \divide\r*\N* \ifnum\r*<2\r*2\fi
\m**\r* \divide\m**2 \l**\r* \advance\l**-\m** \N*\n* \divide\N*\r*
\dt*\d* \divide\dt*\N* \dt*.5\dt* \dt*#1\dt*
\divide\dt*1000\multiply\dt*\magnitude
\k**\r* \multiply\k**\N* \dn*\n* \advance\dn*-\k** \divide\dn*2\relax
\r*\l** \divide\r*2\advance\dn*\r* \advance\N*-1\k**#3\relax
\def\xscale{#5}\def\yscale{#6}\relax
{\angle**0\rotate(#3)}\gmov*(#2,0){\sm*}\advance\k**\dn*
{\angle**0\rotate(\k**)}\gmov*(#2,0){\sl*}\advance\k**-\m**
\advance\l**\m**\loop\dt*-\dt* \d*\d** \advance\d*\dt*
\advance\k**\l**{\angle**0\rotate(\k**)}\rgmov*(\d*,0pt){\sl*}\relax
\advance\N*-1\ifnum\N*>0\repeat\advance\k**\m**
{\angle**0\rotate(\k**)}\gmov*(#2,0){\sl*}\relax
{\angle**0\rotate(#4)}\gmov*(#2,0){\sl*}\def\xscale{1}\def\yscale{1}}}

\def\waveEarc#1(#2)[#3,#4][#5,#6]{{\k**=#3\n*=#4\advance\n*-\k**
\L*=4000sp\L*#2\L* \multiply\L*\n* \multiply\L*\Nhalfperiods
\divide\L*57\N*\L* \divide\N*1000\ifnum\N*=0\N*1\fi
\d**=#2\Lengthunit \d*\d** \divide\d*57\multiply\d*\n*
\r*\n*  \divide\r*\N* \ifnum\r*=0\r*1\fi
\m**\r* \divide\m**2 \l**\r* \advance\l**-\m** \N*\n* \divide\N*\r*
\dt*\d* \divide\dt*\N* \dt*.7\dt* \dt*#1\dt*
\divide\dt*1000\multiply\dt*\magnitude
\k**\r* \multiply\k**\N* \dn*\n* \advance\dn*-\k** \divide\dn*2\relax
\divide\N*4\advance\N*-1\k**#3\def\xscale{#5}\def\yscale{#6}\relax
{\angle**0\rotate(#3)}\gmov*(#2,0){\sm*}\advance\k**\dn*
{\angle**0\rotate(\k**)}\gmov*(#2,0){\sl*}\advance\k**-\m**
\advance\l**\m**\loop\dt*-\dt*
\d*\d** \advance\d*\dt* \dd*\d** \advance\dd*1.3\dt*
\advance\k**\r*{\angle**0\rotate(\k**)}\rgmov*(\d*,0pt){\sl*}\relax
\advance\k**\r*{\angle**0\rotate(\k**)}\rgmov*(\dd*,0pt){\sl*}\relax
\advance\k**\r*{\angle**0\rotate(\k**)}\rgmov*(\d*,0pt){\sl*}\relax
\advance\k**\r*
\advance\N*-1\ifnum\N*>0\repeat\advance\k**\m**
{\angle**0\rotate(\k**)}\gmov*(#2,0){\sl*}\relax
{\angle**0\rotate(#4)}\gmov*(#2,0){\sl*}\def\xscale{1}\def\yscale{1}}}

\newcount\CatcodeOfAtSign
\CatcodeOfAtSign=\the\catcode`\@
\catcode`\@=11
\def\@arc#1[#2][#3]{\rlap{\Lengthunit=#1\Lengthunit
\sm*\l*arc(#2.1914,#3.0381)[#2][#3]\relax
\mov(#2.1914,#3.0381){\l*arc(#2.1622,#3.1084)[#2][#3]}\relax
\mov(#2.3536,#3.1465){\l*arc(#2.1084,#3.1622)[#2][#3]}\relax
\mov(#2.4619,#3.3086){\l*arc(#2.0381,#3.1914)[#2][#3]}}}

\def\dash@arc#1[#2][#3]{\rlap{\Lengthunit=#1\Lengthunit
\d*arc(#2.1914,#3.0381)[#2][#3]\relax
\mov(#2.1914,#3.0381){\d*arc(#2.1622,#3.1084)[#2][#3]}\relax
\mov(#2.3536,#3.1465){\d*arc(#2.1084,#3.1622)[#2][#3]}\relax
\mov(#2.4619,#3.3086){\d*arc(#2.0381,#3.1914)[#2][#3]}}}

\def\wave@arc#1[#2][#3]{\rlap{\Lengthunit=#1\Lengthunit
\w*lin(#2.1914,#3.0381)\relax
\mov(#2.1914,#3.0381){\w*lin(#2.1622,#3.1084)}\relax
\mov(#2.3536,#3.1465){\w*lin(#2.1084,#3.1622)}\relax
\mov(#2.4619,#3.3086){\w*lin(#2.0381,#3.1914)}}}

\def\bezier#1(#2,#3)(#4,#5)(#6,#7){\N*#1\l*\N* \advance\l*\*one
\d* #4\Lengthunit \advance\d* -#2\Lengthunit \multiply\d* \*two
\b* #6\Lengthunit \advance\b* -#2\Lengthunit
\advance\b*-\d* \divide\b*\N*
\d** #5\Lengthunit \advance\d** -#3\Lengthunit \multiply\d** \*two
\b** #7\Lengthunit \advance\b** -#3\Lengthunit
\advance\b** -\d** \divide\b**\N*
\mov(#2,#3){\sm*{\loop\ifnum\m*<\l*
\a*\m*\b* \advance\a*\d* \divide\a*\N* \multiply\a*\m*
\a**\m*\b** \advance\a**\d** \divide\a**\N* \multiply\a**\m*
\rmov*(\a*,\a**){\unhcopy\spl*}\advance\m*\*one\repeat}}}

\catcode`\*=12

\newcount\n@ast
\def\n@ast@#1{\n@ast0\relax\get@ast@#1\end}
\def\get@ast@#1{\ifx#1\end\let\next\relax\else
\ifx#1*\advance\n@ast1\fi\let\next\get@ast@\fi\next}

\newif\if@up \newif\if@dwn
\def\up@down@#1{\@upfalse\@dwnfalse
\if#1u\@uptrue\fi\if#1U\@uptrue\fi\if#1+\@uptrue\fi
\if#1d\@dwntrue\fi\if#1D\@dwntrue\fi\if#1-\@dwntrue\fi}

\def\halfcirc#1(#2)[#3]{{\Lengthunit=#2\Lengthunit\up@down@{#3}\relax
\if@up\mov(0,.5){\@arc[-][-]\@arc[+][-]}\fi
\if@dwn\mov(0,-.5){\@arc[-][+]\@arc[+][+]}\fi
\def\lft{\mov(0,.5){\@arc[-][-]}\mov(0,-.5){\@arc[-][+]}}\relax
\def\rght{\mov(0,.5){\@arc[+][-]}\mov(0,-.5){\@arc[+][+]}}\relax
\if#3l\lft\fi\if#3L\lft\fi\if#3r\rght\fi\if#3R\rght\fi
\n@ast@{#1}\relax
\ifnum\n@ast>0\if@up\shade[+]\fi\if@dwn\shade[-]\fi\fi
\ifnum\n@ast>1\if@up\dshade[+]\fi\if@dwn\dshade[-]\fi\fi}}

\def\halfdashcirc(#1)[#2]{{\Lengthunit=#1\Lengthunit\up@down@{#2}\relax
\if@up\mov(0,.5){\dash@arc[-][-]\dash@arc[+][-]}\fi
\if@dwn\mov(0,-.5){\dash@arc[-][+]\dash@arc[+][+]}\fi
\def\lft{\mov(0,.5){\dash@arc[-][-]}\mov(0,-.5){\dash@arc[-][+]}}\relax
\def\rght{\mov(0,.5){\dash@arc[+][-]}\mov(0,-.5){\dash@arc[+][+]}}\relax
\if#2l\lft\fi\if#2L\lft\fi\if#2r\rght\fi\if#2R\rght\fi}}

\def\halfwavecirc(#1)[#2]{{\Lengthunit=#1\Lengthunit\up@down@{#2}\relax
\if@up\mov(0,.5){\wave@arc[-][-]\wave@arc[+][-]}\fi
\if@dwn\mov(0,-.5){\wave@arc[-][+]\wave@arc[+][+]}\fi
\def\lft{\mov(0,.5){\wave@arc[-][-]}\mov(0,-.5){\wave@arc[-][+]}}\relax
\def\rght{\mov(0,.5){\wave@arc[+][-]}\mov(0,-.5){\wave@arc[+][+]}}\relax
\if#2l\lft\fi\if#2L\lft\fi\if#2r\rght\fi\if#2R\rght\fi}}

\catcode`\*=11

\def\Circle#1(#2){\halfcirc#1(#2)[u]\halfcirc#1(#2)[d]\n@ast@{#1}\relax
\ifnum\n@ast>0\L*=\xscale\Lengthunit
\ifnum\angle**=0\clap{\vrule width#2\L* height.1pt}\else
\L*=#2\L*\L*=.5\L*\special{em:linewidth .001pt}\relax
\rmov*(-\L*,0pt){\sm*}\rmov*(\L*,0pt){\sl*}\relax
\special{em:linewidth \the\linwid*}\fi\fi}

\catcode`\*=12

\def\wavecirc(#1){\halfwavecirc(#1)[u]\halfwavecirc(#1)[d]}

\def\dashcirc(#1){\halfdashcirc(#1)[u]\halfdashcirc(#1)[d]}

\def\xscale{1}
\def\yscale{1}

\def\Ellipse#1(#2)[#3,#4]{\def\xscale{#3}\def\yscale{#4}\relax
\Circle#1(#2)\def\xscale{1}\def\yscale{1}}

\def\dashEllipse(#1)[#2,#3]{\def\xscale{#2}\def\yscale{#3}\relax
\dashcirc(#1)\def\xscale{1}\def\yscale{1}}

\def\waveEllipse(#1)[#2,#3]{\def\xscale{#2}\def\yscale{#3}\relax
\wavecirc(#1)\def\xscale{1}\def\yscale{1}}

\def\halfEllipse#1(#2)[#3][#4,#5]{\def\xscale{#4}\def\yscale{#5}\relax
\halfcirc#1(#2)[#3]\def\xscale{1}\def\yscale{1}}

\def\halfdashEllipse(#1)[#2][#3,#4]{\def\xscale{#3}\def\yscale{#4}\relax
\halfdashcirc(#1)[#2]\def\xscale{1}\def\yscale{1}}

\def\halfwaveEllipse(#1)[#2][#3,#4]{\def\xscale{#3}\def\yscale{#4}\relax
\halfwavecirc(#1)[#2]\def\xscale{1}\def\yscale{1}}

\catcode`\@=\the\CatcodeOfAtSign

\title{\Large The  coupling of fermions to the three-dimensional  noncommutative
$CP^{N-1}$ model: minimal and supersymmetric extensions}

\author{E. A. Asano}
\affiliation{Instituto de Fisica, Universidade de S\~ao Paulo,
Caixa Postal 66318, 05315-970, S\~ao Paulo - SP, Brazil}
\email{asano, mgomes, petrov, alexgr, ajsilva@fma.if.usp.br}
\author{H. O. Girotti}
\affiliation{Instituto de Fisica, Universidade Federal do Rio Grande
do Sul, Caixa Postal 15051, 91501-970 - Porto Alegre, RS, Brazil}
\email{hgirotti@if.ufrgs.br}
\author{M. Gomes}
\author{A. Yu. Petrov}
 \altaffiliation[Also at]{ Department of Theoretical Physics,
Tomsk State Pedagogical University
Tomsk 634041, Russia
(email: petrov@tspu.edu.ru)}
\author{A. G. Rodrigues}
\author{A. J. da Silva}
\affiliation{Instituto de Fisica, Universidade de S\~ao Paulo,
Caixa Postal 66318, 05315-970, S\~ao Paulo - SP, Brazil}
\email{asano, mgomes, petrov, alexgr, ajsilva@fma.if.usp.br}

\begin{abstract}

We consider the coupling of fermions to the three-dimensional
noncommutative $CP^{N-1}$ model.  In the case of minimal coupling,
although the infrared behavior of the gauge sector is improved, there
are dangerous (quadratic) infrared divergences in the corrections to the two
point vertex function of the scalar field.  However, using superfield
techniques we prove that the supersymmetric version of this model with
``antisymmetrized'' coupling of the Lagrange multiplier field is
renormalizable up to the first order in $\frac{1}{N}$. 
The auxiliary spinor gauge field acquires a nontrivial (nonlocal) dynamics with
a generation of Maxwell and Chern-Simons noncommutative terms in the effective action. Up to the $1/N$ order all divergences are only logarithimic so that
the model is free from nonintegrable infrared singularities.

\end{abstract}

\maketitle
\newpage

\section{Introduction}
The renormalization problem is a central issue for the perturbative
consistency of noncommutative (NC) field theories. This is of course
true for any field theory but for the noncommutative setting
renormalization becomes more stringent due to an unusual mixture of
scales.  In fact, a characteristic phenomena in such theories is the
well known ultraviolet/infrared (UV/IR) mixing which, being the source
of nonintegrable IR divergences \cite{Minw} (for a review see
\cite{Nekr}), destroys most of the perturbative schemes. It is
therefore very important to find renormalizable noncommutative field
theories free from the mentioned infrared divergences.  We have
recently proved that, at least up to next to leading order of $1/N$,
this requirement is satisfied by the ($2+1$) dimensional
noncommutative version of the $CP^{N-1}$ model if the basic field
transforms in accord with the fundamental representation of the gauge
group~\cite{Gomes1}. For the same model, we also investigated the
situation where the basic field belongs to the adjoint representation.
In contrast with the fundamental representation, we found that for the
adjoint representation infrared divergences associated to nonplanar
graphs are present. These infrared divergences indicate the breakdown
of the model at higher orders of $1/N$.  Our previous experience with
the noncommutative versions of the four dimensional Wess-Zumino model
\cite{WZ} as well as with the (2+1) dimensional supersymmetric
nonlinear sigma model \cite{sig}, suggests that the overall behavior
of the theory may be improved if fermions are included.  In this paper
we will investigate such possibility by coupling fermions to the gauge
field either minimally or in a supersymmetric fashion.  Of course,
even in the case of minimal coupling, the fermionic field and its
bosonic counterpart must belong to the same representation.

Very interesting results emerge from our analysis.  As we shall prove,
due to the induction of a Chern-Simons term, the gauge
potential becomes much less singular. However, in the case of minimal
coupling, in spite of the general smootheness of the gauge potential,
the radiative
corrections to the self energy of the scalar field
are still plagued by
 nonintegrable infrared singularities. To evade this
problem we then consider a supersymmetric extension of the model. This
is done through the use of powerful superfield techniques \cite{BK0,SGRS},
which enable us to demonstrate the absence of the dangerous UV/IR
mixing up to order $1/N$.  

Our work is organized as follows. In Sec. \ref{fermion} the inclusion
of fermion fields minimally coupled to the gauge field is examined. In
Sec. \ref{susy} the superfield formulation is introduced, we fix the
notation to be employed and determine the propagators for the relevant
fields. In Sec. \ref{sec4} we prove that the self-energy corrections
of the scalar superfield are free from dangerous UV/IR mixing and in
Sec. \ref{sec5} give a general argument for the absence of these
singularities in all Green functions up to $1/N$ order.  A general
overview of our results and the conclusions are contained in
Sec. \ref{sec6}.

 \section{Minimal coupling of fermions to the $CP^{N-1}$ model}
\label{fermion}

Assuming that the
fermions have the same mass as their bosonic counterpart, the action
associated to the model reads (for discussions on the commutative $CP^{N-1}$
model see \cite{Vec,Macfa,Aoy,Duer,ArA})

\begin{equation}
\int d^3 x {\mathcal{L}}=\int d^3 x [-(D_{\mu }\varphi )^{\dag }*D^{\mu }
\varphi
-m^{2}\varphi ^{\dag }*\varphi -\bar{\psi }*\gamma _{\mu }D^{\mu }\psi-m\bar{\psi }*\psi
+{\cal L}_{\lambda}]
,\end{equation}

\noindent
 where $\varphi_a$ and $\psi _{a}$, $a=1,\ldots ,N$ are scalar and
two-component Dirac fields, respectively. They transform according to
either the left fundamental or the adjoint representation of the gauge
group. To keep uniformity throughout this work, we shall use the
metric $g_{11}=g_{22}=-g_{00}=1$ and the Dirac matrices to be employed
in this section are $\gamma ^{0}=i\sigma ^{3},\, \gamma ^{1}=\sigma
^{1}$ and $\gamma ^{2}=\sigma ^{2}$, where the $\sigma $'s are the
Pauli matrices).  The covariant derivative of the basic
fields is $D_\mu\chi=\partial _{\mu }\chi +iA_{\mu }\ast \chi$, for
$\chi=\varphi, \psi$ in the left fundamental representation, whereas $D_{\mu }\chi
=\partial _{\mu }\chi +iA_{\mu }\ast \chi -i\chi \ast A_{\mu }$, for
$\chi=\varphi, \psi$ in the adjoint representation. ${\cal L}_\lambda$
is the interaction Lagrangian which enforces a basic constraint
for the $\varphi_a$ fields; its possible forms will be
given shortly. 
Besides, to evade
unitarity problems, throughout this work we consider only space-space
noncommutativity.

\subsection{The bosonic model}
We begin by recalling some basic results of the pure $CP^{N-1}$ model, i.e. without fermions~\cite{Gomes1}:


(1) For the left fundamental representation case, with ${\cal
  L}_\lambda=\lambda * (\varphi*\varphi^\dag -\frac{N}{g})$, the two
point vertex functions of the gauge and $\lambda$  fields are respectively:

\begin{equation}
F_{b}^{\mu\nu}(p)=-\frac{iN}{8\pi }(g^{\mu \nu }p^2-p^{\mu }p^{\nu })
\int
_{0}^{1}dx\frac{(1-2x)^{2}}{M(x)} ,\label{10}
\end{equation}

\noindent
and

\begin{equation}
F(p)=N\int \frac{d^{3}k}{(2\pi )^{3}}\frac{1}{(k+p)^{2}+m^{2}}
\frac{1}{k^{2}+m^{2}}= \frac{iN}{8\pi}\int_{0}^{1} dx \frac{1}{M(x)},\label{9}
\end{equation}

\noindent where $M(x)={[m^{2}+p^{2}x(1-x)]^{1/2}}$. Furthermore, the mixed two point vertex function $F_\mu$ of the $A_\mu$ and $\lambda$ fields vanishes.

(2) For the adjoint representation there are two cases that have to be distinguished:

(2a) The part of the interaction Lagrangian which contains $\lambda$
is ${\cal L}_\lambda=\lambda*[\varphi,\varphi^\dag]_*$.  Here also the mixed two point vertex function $F_\mu$ vanishes.

The two
point vertex function of the $A_\mu$ field is 

\begin{eqnarray}
F_{b}^{\mu \nu }(p)=-\frac{iN}{4\pi } &  &  \left\{ (g^{\mu \nu }
p^2-
p^{\mu }p^{\nu }) \int _{0}^{1}dx\frac{(1-2x)^2}{M}(1-{\textrm{e}}^{-M\sqrt{\tilde{p}^{2}}})
\right.\nonumber \\
 &  & \left.+4\frac{\tilde{p}^{\mu }\tilde{p}^{\nu }}{\tilde{p}^{2}}
\int _{0}^{1}dx(\frac{1}{\sqrt{\tilde{p}^{2}}}+M){\textrm{e}}^{-M
\sqrt{\tilde{p}^{2}}}\right\} ,\label{n11a}
\end{eqnarray}

 \noindent 
in which $\tilde p_\mu
=\theta_{\mu\nu} p^\nu$ and
$\theta_{\mu\nu}$ is the constant antisymmetric matrix characterizing the
noncommutativity of the underlying space.  Notice that the above result is transversal
but possesses an infrared singularity at $\tilde{p}=0$. 

The two point vertex function of the $\lambda$ field is modified to

\begin{equation}
2{F(p)}+ F_{NP}(p)\equiv\frac{iN}{4\pi} f(p),\label{50}
\end{equation}

\noindent
where $F$ was given in (\ref{9}) and the nonplanar part 
$F_{NP}$ is

\begin{equation}  
 F_{NP}(p)=-\frac{i N}{4\pi}\int_{0}^{1}dx\, \frac{{\rm e}^{-M
\sqrt{{\tilde p}^2}}}{ M}.
\end{equation}

The function $f(p)$ is explicitly given by

\begin{equation}\label{19}
f(p) = \int_{0}^{1}dx\frac{1-{\textrm{e}}^{-M\sqrt{{\tilde{p}}^{2}}}}{M}
\approx\left \{ \begin{array}{l@{\hspace{20 pt}}rl}
\sqrt{{\tilde p}^2} & \mbox{for} & p\rightarrow 0, \\[16pt]
{\pi}/{\sqrt{{ p}^2}} & \mbox{for} & p^2 \gg m^2.
\end{array}
\right.
\end{equation}

For future use it is convenient to identify

\begin{equation}
f(p) = -16 \pi i \int \frac{d^3 k}{(2\pi)^3}\, I(k,p),\label{gel}
\end{equation}

\noindent
where 

\begin{eqnarray}
I(k,p)=\frac{\sin^2(k\wedge p)}{(k^2+m^2)((k+p)^2+m^2)}
\end{eqnarray}
 
\noindent
and $k\wedge p=\frac12 k\cdot {\tilde p}$.

(2b) The interaction Lagrangian ${\cal L}_\lambda$ has the same form as in the 
case of the left fundamental representation. The two point vertex functions of the $A_\nu$ and
$\lambda$ fields are still given by (\ref{n11a}) and (\ref{9})
but now there
exists a nonvanishing mixed two point vertex function

\begin{eqnarray}
F _\mu(p)&=&N\int \frac{d^{3}k}{(2\pi )^{3}}\, 
\frac{(2k+p)_{\mu }}{(k^{2}+m^{2})[(k+p)^{2}+m^{2}]}{
\textrm{e}}^{-i2k\wedge p}\nonumber \\
&=&\frac{N\tilde{p_{\mu }}}{4\pi 
\sqrt{{\tilde{p}}^{2}}}
\int_{0}^{1} dx {\rm e}^{-M\sqrt{{\tilde p}^2}}\equiv \frac{N g(p)}{4\pi}\tilde p_\mu.
\label{49}
\end{eqnarray}

\subsection{Including fermions}

 Due to the inclusion of fermionic fields, the two point vertex
function of the $A_{\mu }$ field receives a new contribution:

\begin{equation}
F^{\mu \nu }_f(p)=-N \int \frac{d^{3}k}{(2\pi )^{3}}{\textrm{Tr}}
[\gamma ^{\nu }\frac{i}{-i\not \! k+m}\gamma ^{\mu }\frac{i}{-i(\not \! k+
\not \! p)+m}]J(k,p),\label{NCfermionloop}
\end{equation}

\noindent 
where $J(k,p)$ is either equal to one or to $4\sin ^{2}{(k\wedge p)}$
for the left fundamental or the adjoint representations, respectively.
In (\ref{NCfermionloop}) the subscript $f$ was used to designate
the fermionic part. After some standard manipulations, we arrive at

\begin{eqnarray}
F^{\mu \nu }_f(p) & = & -2N\int _{0}^{1}dx\int \frac{d^{3}k}{(2\pi )^{3}}J(k,p)
\nonumber \\
 & \phantom a & \times\frac{2k^{\mu }k^{\nu }-2p^{\mu }p^{\nu }x(1-x)-
g^{\mu \nu }[k^{2}-p^{2}x(1-x)+m^{2}]+im\epsilon ^{\mu \nu \rho }p_{\rho }}
{[k^{2}+p^{2}x(1-x)
+m^{2}]^{2}}.
\end{eqnarray}

For the left fundamental representation this produces  the well
known result \cite{Schonfeld,Jackiw}

\begin{equation}
F^{\mu\nu}_f(p)=-\frac{Ni}{2\pi }(g^{\mu \nu }p^{2}-p^{\mu }p^{\nu })\int _{0}^{1}dx
\frac{x(1-x)}{M}+\frac{mN}{4\pi }\epsilon ^{\mu \nu \rho }p_{\rho }
\int _{0}^{1}dx\frac{1}{M}.\label{left}\end{equation}

For the adjoint representation, the  use of $\sin ^{2}{(k\wedge
p)}=\frac{1-\cos (2{k\wedge p})}2$  leads to a planar contribution
equal to twice the above result.  The nonplanar
contribution (which contains the factor $\cos (2{k\wedge p})$) gives

\begin{eqnarray}
F^{\mu \nu }_{NPf}(p) & = & \frac{iN}{\pi }(g^{\mu \nu }p^{2}-p^{\mu }p^{\nu })
\int _{0}^{1}dx\frac{x(1-x)}{M}{\rm e}^{-M\sqrt{\tilde{p}^{2}}}
+\frac{iN}{\pi }\frac{\tilde{p}^{\mu }\tilde{p}^{\nu }}{\tilde{p}^{2}}
\int _{0}^{1}dx(M+\frac{1}{\sqrt{\tilde{p}^{2}}}){\rm e}^{-M
\sqrt{\tilde{p}^{2}}}\nonumber \\
 &  & -\frac{mN}{2\pi }\epsilon ^{\mu \nu \rho }p_{\rho }\int _{0}^{1}dx
\frac{1}{M}e^{-M\sqrt{\tilde{p}^{2}}}.\label{n11b}
\end{eqnarray}

Thus, by adding the contributions from the bosonic and fermionic loops we get 
the total two
point vertex function of the gauge field as being:

1. For the left fundamental representation  (sum of Eqs. (\ref{10}) plus
(\ref{left}) )

\begin{equation}
F^{\mu \nu }(p)=\frac{-iN}{8\pi }[(g^{\mu \nu }p^{2}-p^{\mu }p^{\nu })
+{2im }\epsilon ^{\mu \nu \rho }p_{\rho } ]\int_{0}^{1} \frac{dx}{M}.
\label{fund}
\end{equation}

2. For the adjoint representation (sum of Eqs. (\ref{n11a}) plus twice (\ref{left}) 
plus (\ref{n11b}))

\begin{equation}
F^{\mu \nu }(p)=\frac{-iN}{4\pi }f(p)[(g^{\mu \nu }p^{2}-p^{\mu }p^{\nu })
+{2im }\epsilon ^{\mu \nu \rho }p_{\rho } ].
\label{adj}\end{equation}

\noindent
As can be seen, $F^{\mu \nu }(p)$ behaves smoothly as $p$ tends to
zero. Notice the presence of terms proportional to
$\epsilon^{\mu\nu\rho} $ in Eqs.  (\ref{left}) and (\ref{adj}),
which in the effective action correspond to the bilinear part of the
noncommutative Chern-Simons term.  From now on we will restrict
our considerations to the adjoint representation. 

For the case (2a) the propagator for the $\lambda$ field is $\Delta(p)=\frac{4\pi i}{ 
N f(p)}$ and   the propagator for the gauge field in the Landau gauge is

\begin{equation}
\Delta _{\mu \nu }(p)=\frac{-4\pi i}{Nf(p)(p^{2}+4m^{2})}
[(g_{\mu \nu }-\frac{p_{\mu }\, p_{\nu }}{p^{2}})-\frac{2im}{ p^{2}}
\epsilon _{\mu \nu \rho }p^{\rho }].\label{adj1}
\end{equation}

For the case (2b), due to the nonvanishing mixed two point vertex function of the $\lambda$ and $A_\mu$
fields, the computation of the gauge field  propagator is much more involved than in the
previous case. We find (also in the Landau gauge)

\begin{eqnarray}
\Delta_{\mu\nu}(p) &=& A_1 (g_{\mu\nu} - \frac{p_\mu p_\nu}{p^2}) +
A_2 {\tilde p}_\mu 
{\tilde p}_\nu + 
A_3 {\bar p}_\mu {\bar p}_\nu + A_4( {\tilde p}_\mu{\bar p}_\nu-
{\tilde p}_\nu{\bar p}_\mu) + A_5 \epsilon_{\mu\nu\rho} p^\rho\nonumber \\
&=& (A_1- p^2 {\tilde p}^2 A_3)(g_{\mu\nu} - \frac{p_\mu p_\nu}{p^2})+
(A_2+ p^2 A_3){\tilde p}_\mu  {\tilde p}_\nu + (A_5+ {\tilde p}^2 A_4)
\epsilon_{\mu\nu\alpha} p^\alpha,
\end{eqnarray}

\noindent
where ${\bar p}_\alpha\equiv \epsilon_{\alpha\beta\gamma} p^\beta
 {\tilde p}^\gamma$ and the coefficients $A_i$, $i=1,\ldots,5$, are functions
of $p$:

\begin{eqnarray}
A_1&=&-i\frac{4\pi }{N}\frac{1}{f(p)} \frac{1}{p^2+ 4 m^2}, \quad\qquad 
A_2 =
\frac{4\pi }{N}\frac{g^2(p)}{h(p)} \frac{1}{p^2+ 4 m^2},\\
A_3&=& \frac{4 \pi  }{N}\frac{4 m^2g^2(p)}{h(p)(p^2)^2} \frac{1}{p^2+ 4 m^2}, \quad \quad
A_4= i\frac{4 \pi}{N} \frac{2m g^2(p)}{h(p)p^2  }\frac{1}{p^2+ 4 m^2},
\end{eqnarray}

\noindent
and
\begin{equation}
A_5 =- \frac{4\pi }{N}\frac{2m}{f(p)}\frac{1}{p^2(p^2+ 4 m^2)},
\end{equation}
\noindent
where
\begin{equation}
h(p)= - if(p)  \left [ {\tilde p}^2 g^2(p)+ f^2(p)   
(p^2+ 4 m^2) \right ].
\end{equation}

For large momenta this propagator coincides  with that in Eq. (\ref{adj1}),
since $g(p)$ exponentially decreases or strongly oscillates in that limit.

Notice that in both situations the gauge propagator is much less singular than in the pure
$CP^{N-1}$ model. This smoothness of the infrared behavior comes as a
direct effect of the generation of the Chern-Simons term which
provided the displacement from the origin of the usual ($p^2=0$)
singularity.

For reference we also quote the expressions for the $\lambda$ and mixed
($\lambda,  A_\nu$) propagators

\begin{equation}
\Delta(p) =\frac{4\pi}N \frac{f^2(p)}{h(p)}(p^2+ 4 m^2), \qquad \Delta_\nu(p)
= -\frac{4\pi}N \frac{f(p)g(p)}{h(p)} ( i {\tilde p}_\nu + \frac{2 m{\bar p}_\nu}{p^2}).
\end{equation}

Although we have improved the infrared behavior of the $A_{\mu }$
propagator we still get trouble with the radiative corrections to the
propagator for the $\varphi$ field. In fact, whereas graph
\ref{Fig5}$b$ is finite (in the Landau gauge), a direct calculation shows that the nonplanar
parts of the graphs of Figs. \ref{Fig5}$a$ and \ref{Fig5}$c$ are
infrared quadratically divergent. Up to the $1/N$ order they are the
only infrared divergent diagrams contributing to the self energy of
$\varphi$ field. The sum of their nonplanar parts does not vanish and
therefore, at higher orders, leads to a breakdown of the $1/N$
expansion \cite{Gomes1}. To overcome this problem a further extension of the model
is needed. This will be the subject of the following sections where we
discuss a  supersymmetric extension of the noncommutative $CP^{N-1}$ model.

\section{The noncommutative supersymmetric $CP^{N-1}$ model}
\label{susy}

In the adjoint representation the noncommutative superfield
generalization of the $CP^{N-1}$ model is described by (see also
\cite{saito,oh} for  supersymmetric extensions of its commutative
counterpart)
\begin{eqnarray}
\label{acti}
S&=&-\int d^5 z \Big[\frac12(D^{\a}\bar{\phi}_a+i[\bar{\phi}_a,A^{\a}]_*)*
(D_{\a}\phi_a-i[A_{\a},\phi_a]_*)+m\bar{\phi}_a\phi_a\nonumber\\&\phantom a&+
\eta*(a[\bar{\phi}_a,\phi_a]_*+ b\{\bar{\phi}_a,\phi_a\}_*)-\eta\frac{Nb}{g}
\Big],
\end{eqnarray}

\noindent
where $\phi_a$ with $a=1\ldots N$ is a set of scalar (super) fields,
$\bar{\phi}_a$ are their complex conjugated ones, $A_{\a}$ is a
two-component spinor gauge field and $\eta$ is a Lagrange multiplier
superfield which implements the constraint
$\{\bar{\phi}_a,\phi_a\}_*\equiv\bar{\phi}_a * \phi_a +{\phi}_a *\bar
\phi_a =\frac{N}{g}$; $a$ and $b$ are parameters which control the two possible
orderings of the trilinear term containing the $\eta$, $\phi$ and $\bar \phi$
fields.  Hereafter, we employ the same notation and
definitions of \cite{SGRS} (see also a description of the
three-dimensional superfield approach in \cite{RR}). Concisely,
$\l^2\equiv\frac12 \l^{\a}\l_{\a}=\hf C^{\a\b}\l_{\b}\l_{\a}$ for any
spinor $\l^{\a}$ (and $D^2=\hf D^{\a}D_{\a}$), with
$C_{\a\b}=-C^{\alpha\beta}$ an antisymmetric matrix normalized as
$C_{12}=-i$, $\psi_{\a}=\psi^{\b}C_{\b\a}$ and
$\psi^{\a}=C^{\a\b}\psi_{\b}$. The Dirac matrices with both spinor
indices upstairs are $\gamma^m=({\bf 1},\s^3,-\s^1)$ and satisfy
$\{\gamma^m,\gamma^n\}=2g^{mn}{\bf 1}$.

The above action is invariant under the infinitesimal supergauge 
transformation:

\begin{eqnarray}
\delta \f= i [K,\,\f]_*\, , \qquad \delta \eta= i [K,\,\eta]_*,\qquad
\delta A_\alpha= D_\alpha K +i [K, A_\alpha]_*,
\end{eqnarray} 

\noindent
where $K$ is the scalar superfield gauge parameter. We will consider
two cases, namely, the {\it commutator} case when $a=1$ and $b=0$ and
the {\it anticommutator} case when $a=0$ and $b=1$. Notice that
dynamical generation of mass only occurs in the anticommutator case
(the analysis is entirely similar to the one in \cite{sig}). 
In the sequel
we will be explicitly analyzing the commutator case but we will
also comment on the other possibility.

As it is well known, charge conjugation (and parity) are in general
broken for noncommutative field theories \cite{Sheikh}.  Notice
however, that for the commutator case the above action is invariant
under the ``charge conjugation'' transformation $\phi \leftrightarrow
\bar \phi$, $A_\alpha \rightarrow A_\alpha$, and  $\eta \rightarrow -\eta$
and, as a consequence, the ``mixed propagator''
$<\eta A_\alpha>$ vanishes. This last conclusion depends crucially on
the way in which the $\eta$ and $\phi$ fields are coupled. Had we used
an anticommutator in the term multiplying the $\eta$ field, then
$\eta$ would be even under charge conjugation and the mixed propagator
would not vanish.
For the commutator case, an equivalent and useful form for the above action is

\begin{eqnarray}
S&=&\int d^5 z
\Big[\bar{\phi}_a(D^2-m)\phi_a-\frac{i}{2} ([\bar{\phi}_a,A^{\a}]_**D_{\a}\phi_a-
D^{\a}\bar{\phi}_a*[A_{\a},\phi_a]_*)
\nonumber\\&\phantom a &
-\hf [\bar{\phi}_a,A^{\a}]_**[A_{\a},\phi_a]_*-\eta*[\bar{\phi}_a,\phi_a]_*
\Big].
\end{eqnarray}

As in the pure $CP^{N-1}$ model, at the classical level only the scalar fields are dynamical but
quantum corrections may provide effective dynamics for the other fields
(compare also with \cite{sig}). All fields belong to the adjoint
representation of the gauge group which explains the commutators in
the terms involving the gauge field; these commutators cause sine
factors in the corresponding vertices of the Feynman graphs. 
Using that $(D^2)^2=\Box$  and $(D^2+m)(-D^2+m)=-\Box+m^2$
we obtain the  free propagator for the scalar fields 

\begin{equation}
<T\bar{\phi}_a(z_1)\phi_b(z_2)>=i\delta_{ab}\frac{D^2+m}{\Box-m^2}
\delta^5(z_1-z_2),
\end{equation}

\noindent
which, in   momentum space reads

\begin{eqnarray}
<T\bar{\phi}_a(k_1,\theta_1)\phi_b(k_2,\theta_2)>= (2\pi)^3\delta^3(k_1+k_2) 
<\bar{\phi}_a(k_1,\theta_1)\phi_b(-k_1,\theta_2)>,
\end{eqnarray}

\noindent
where
\begin{equation}
<\bar{\phi}_a(k,\theta_1)\phi_b(-k,\theta_2)>=
-i
\delta_{ab}\frac{D^2+m}{k^2+m^2}\delta_{12},
\end{equation}

\noindent
with $\delta_{12}\equiv\delta^2(\theta_1 - \theta_2)$.

Let us now  obtain the effective propagators for the $\eta$ and $A^{\a}$ fields.
First we turn to the  $\eta$ field. The effective propagator is
generated by the supergraph of Fig. \ref{Fig3}.
The contribution of this graph to the effective action $S_2=\int\frac 
{d^3p}{(2\pi)^3}\,S_2(p)$ is

\begin{eqnarray}
iS_2(p)&=& 2N\int d^2\theta_1 d^2\theta_2\int\frac{d^3k}{(2\pi)^3} I(k,p)\nonumber\\&\times&
(D^2+m)\delta_{12}(D^2+m)\delta_{12}\eta(-p,\theta_1)\eta(p,\theta_2).
\end{eqnarray}

\noindent
Performing  D-algebra transformations in a way analogous to the derivation of
the effective propagator for $\S$ field in \cite{sig}, we get

\begin{equation}
S_2(p)=\frac{N}{8\pi} f(p) \int d^2 \theta\,
\eta(-p,\q)(D^2+2m)\eta(p,\q),\label{1}
\end{equation}

\noindent
where $f(p)$ was defined in Eq. (\ref{19}). From this expression
we can obtain the propagator for the $\eta$ field:

\begin{eqnarray}
\label{peta}
<\eta(p,\theta_1)\eta(-p,\theta_2)>=
-\frac{4\pi i}{N}\frac{D^2-2m}{f(p)({p^2+4m^2})}\delta_{12}.
\end{eqnarray}

This propagator is linearly divergent for small $p$, since 
in this limit $f(p)\approx \sqrt{{\tilde p}^2}$.
 However, this divergence does not bring difficulties since, for
zero momentum, the radiative corrections to the two point vertex
function of the $\eta$ field will also vanish (as a consequence of the
sine factors at the vertices). On the other hand, for high momenta the
nonplanar contribution in Eq. (\ref{peta})  rapidly decreases. 
 Therefore, when  analyzing the ultraviolet behavior
of Feynman amplitudes we can take just the asymptotic behavior of the
planar part which furnishes

\begin{eqnarray}
<\eta(p,\theta_1)\eta(-p,\theta_2)>\approx
-\frac{4 i}{N}\frac{D^2-2m}{\sqrt{p^2}}\delta_{12}.
\end{eqnarray}

 Next, we turn to the effective propagator of the spinor field $A_{\a}$.
It is formed by the two contributions shown in Fig. \ref{Fig4}.
The first graph, depicted in  Fig \ref{Fig4}$a$, gives the following contribution:

\begin{eqnarray}
iS_{3a}(p)&=&-\int d^2\theta_1 d^2\theta_2 \int\frac{d^3k}{(2\pi)^3}A^{\a}(-p,
\theta_1)
A^{\b}(p,\theta_2)\sin^2(k\wedge p)
\\&&\times\Big[
D_{\a 1}<\phi_a(-k,\theta_1)\bar{\phi}_b(k,\theta_2)>(<\bar{\phi}_a(k+p,\theta_1)\f_b(-k-p,\theta_2)>\der_{\phantom a \b 2})\nonumber\\
&&-
(D_{\a 1}<\phi_a(-k,\theta_1)\bar{\phi}_b(k,\theta_2)>\der_{\phantom a \b 2})
<\bar{\phi}_a(k+p,\theta_1)\f_b(-k-p,\theta_2)>
\Big],\nonumber
\end{eqnarray}

\noindent
where the notation $D_{\gamma i}$ was used to indicate that the supercovariant derivative is applied to the field whose Grassmanian argument is $\theta_i$.
Taking into account the explicit form of the propagators, we have 

\begin{eqnarray}
iS_{3a}(p)&=& N \int d^2\theta_1 d^2\theta_2 \int\frac{d^3k}{(2\pi)^3}A^{\a}
(-p,\theta_1)
A^{\b}(p,\theta_2)\sin^2(k\wedge p)\nonumber\\&\times&\Big[
\frac{D_{\a 1}(D^2_1+m)}{k^2+m^2}\delta_{12}
\frac{(D^2_1+m)D_{\b 2}}{(k+p)^2+m^2}\delta_{12}
\nonumber\\&-&
\frac{D_{\a 1}(D^2_1+m)D_{\b 2}}{k^2+m^2}\delta_{12}
\frac{D^2_1+m}{(k+p)^2+m^2}\delta_{12}
\Big].
\end{eqnarray}

\noindent
Integrating by parts some of the spinor derivatives and by using the identity~$D_{\b 2}(k,\theta_2)\delta_{12}=-D_{\b 1}(-k,\theta_1)\delta_{12}$
we arrive at 

\begin{eqnarray}
\label{expr}
iS_{3a}(p)&=&N \int d^2\theta_1 d^2\theta_2 \int\frac{d^3k}{(2\pi)^3}
I(k,p)
\nonumber\\&\times&\Big[
2(D^2_1+m)\delta_{12}
D_{\a 1}(D^2_1+m)D_{\b 1}\delta_{12}
A^{\a}(-p,\theta_1) A^{\b}(p,\theta_2)
\nonumber\\&+&
(D^2_1+m)\delta_{12} (D^2_1+m)D_{\b 1}\delta_{12}
(D^{\a}A_{\a})(-p,\theta_1) A^{\b}(p,\theta_2)
\Big].
\end{eqnarray}

\noindent
It is convenient to separate $S_{3a}$ into two parts,
$S_{3a}=S_{3a}^{(1)}+S_{3a}^{(2)}$, where $iS^{(1)}_{3a}$ and
$iS_{3a}^{(2)}$ are associated to the two terms in the large brackets
of (\ref{expr}).  Consider first $iS^{(1)}_{3a}$ which, after
transporting $D^2$ from one of the propagators to the other factors,
becomes

\begin{eqnarray}
iS^{(1)}_{3a}(p)&=&N \int d^2\theta_1 d^2\theta_2 \int\frac{d^3k}{(2\pi)^3} I(k,p)
\nonumber\\&\times&\Big[
2m\delta_{12}
D_{\a 1}(D^2_1+m)D_{\b 1}\delta_{12}
A^{\a}(-p,\theta_1) A^{\b}(p,\theta_2)
\nonumber\\&+&2\delta_{12}D^2_1\Big( D_{\a 1}(D^2_1+m)D_{\b 1}\delta_{12}
A^{\a}(-p,\theta_1)\Big) A^{\b}(p,\theta_2)
\Big].
\end{eqnarray}

\noindent
Now we employ the identity $\{D_{\a 1}, D^2_1\}=0$ \cite{SGRS} which leads to

\begin{eqnarray}
iS^{(1)}_{3a}(p)&=&N \int d^2\theta_1 d^2\theta_2 \int\frac{d^3k}
{(2\pi)^3} I(k,p)
\\&\times&\Big[
2\delta_{12}(k^2+m^2)D_{\a 1}D_{\b 1}
\delta_{12}A^{\a}(-p,\theta_1) A^{\b}(p,\theta_2)\nonumber\\&+&
2\delta_{12}(-D^2_1+m)D_{\a 1}D_{\b 1}\delta_{12}
(D^2A^{\a}(-p,\theta_1)) A^{\b}(p,\theta_2)
\Big].\nonumber
\end{eqnarray}

\noindent
The use of  the relationship 

\begin{eqnarray}
D_{\a}(-k, \theta_1)D_{\b}(-k, \theta_1)=k_{\a\b}-C_{\a\b}D^2(-k, \theta_1)\label{idsg}
\end{eqnarray}

\noindent
now provides

\begin{eqnarray}
iS^{(1)}_{3a}(p)&=&2N \int d^2\theta_1 d^2\theta_2 \int\frac{d^3k}{(2\pi)^3} 
I(k,p)
\nonumber\\&\times&\Big[
\delta_{12}(k^2+m^2)(k_{\a\b}-C_{\a\b}D^2)
\delta_{12}A^{\a}(-p,\theta_1) A^{\b}(p,\theta_2)
\nonumber\\&+&
\delta_{12}(-D^2+m)(k_{\a\b}-C_{\a\b}D^2)\delta_{12}
(D^2A^{\a}(-p,\theta_1)) A^{\b}(p,\theta_2)
\Big].
\end{eqnarray}

The only terms giving non-zero contributions are those containing just
one $D^2$ since $\delta_{12}D^2\delta_{12}=\delta_{12}$.  Indeed, by
employing this identity and after integrating over $\theta_2$ with the
help of the delta function, we obtain

\begin{eqnarray}\label{1a}
iS^{(1)}_{3a}(p)&=&-2N \int d^2\q  \int\frac{d^3k}{(2\pi)^3} I(k,p)
\\&\times&\Big[
(k^2+m^2)C_{\a\b} A^{\a}(-p,\q)A^{\b}(p,\q)
+(k_{\a\b}+mC_{\a\b})(D^2A^{\a}(-p,\q)) A^{\b}(p,\q)
\Big].\nonumber
\end{eqnarray}

\noindent
The second term of (\ref{expr}) is

\begin{eqnarray}
iS^{(2)}_{3a}(p)&=&N \int d^2\theta_1 d^2\theta_2 \int\frac{d^3k}{(2\pi)^3}
I(k,p)
\nonumber\\&\times&\Big[
(D^2_1+m)\delta_{12} (D^2_1+m)D_{\b 1}\delta_{12}
(D^{\a}A_{\a})(-p,\theta_1) A^{\b}(p,\theta_2)
\Big].
\end{eqnarray}

\noindent
In this expression we must keep only the  term proportional to 
$D^2_1\delta_{12}(D^2_1+m)D_{\b1}\delta_{12}$ (the remaining part is a
trace of an  odd number of derivatives and therefore vanishes).
Thus, after manipulations similar to those done for 
$S^{(1)}_{3a}$, we find

\begin{eqnarray}
iS^{(2)}_{3a}(p)=-N\int d^2\q  \int\frac{d^3k}{(2\pi)^3} I(k,p)
\Big[D^{\gamma}D^{\a}A_{\a}(-p,\q)
(k_{\gamma\b}+mC_{\gamma\b})A^{\b}(p,\q)
\Big].\label{2}
\end{eqnarray}
\noindent
By adding (\ref{1a}) and (\ref{2}) we can write the total contribution from  
Fig. \ref{Fig4}$a$  as

\begin{eqnarray}
\label{s1}
iS_{3a}(p)&=&-2N \int d^2\q \int \frac{d^3k}{(2\pi)^3} 
I(k,p)
\nonumber\\&\times&\Big[
(k^2+m^2)C_{\a\b} A^{\a}(-p,\q)A^{\b}(p,\q)
+(k_{\a\b}+mC_{\a\b})(D^2A^{\a}(-p,\q)) A^{\b}(p,\q)
\nonumber\\&+&\hf D^{\gamma}D^{\a}A_{\a}
(-p,\theta)(k_{\gamma\b}+mC_{\gamma\b})A^{\b}(p,\q)
\Big].
\end{eqnarray}

\noindent
The algebraic manipulations for the graph  \ref{Fig4}$b$ are considerably 
simpler and yield
 
\begin{eqnarray}
\label{s2}
iS_{3b}(p)&=&2N\int\frac{d^3k}{(2\pi)^3}\frac{\sin^2(k\wedge p)}{(k+p)^2+m^2}
C_{\a\b} A^{\a}(-p,\q)A^{\b}(p,\q).
\end{eqnarray}

\noindent
The complete two point vertex  function for the $A_\alpha$ field is the sum of (\ref{s1}) and (\ref{s2}) and
therefore reads

\begin{eqnarray}
\label{stot}
iS_3(p)&=&-2N \int d^2\q \int \frac{d^3k}{(2\pi)^3} 
I(k,p)
\nonumber\\&\times&
(k_{\gamma\beta}+mC_{\gamma\beta})\Big[(D^2A^{\gamma}(-p,\q)) A^{\b}(p,\q)
+\hf D^{\gamma}D^{\a}A_{\a}(-p,\q) A^{\b}(p,\q)
\Big].
\end{eqnarray}

\noindent
Observe that the dangerous linear divergences (as well as the
logarithimic ones) present in $S_{3a}$ and $S_{3b}$ were canceled in
the above result (compare with \cite{qed4}). As a consequence the free
two point vertex function of the gauge field does not present UV/IR
mixing. Furthermore, notice that the graphs in the Figs. \ref{Fig3}
and \ref{Fig4} cannot occur as subgraphs of more complicated diagrams,
i.e. they are ``illegal'' subgraphs, since they have already been
taken into account to construct the propagators for the $A^{\alpha}$
and $\eta$ fields.

The two point vertex function (\ref{stot})
allows us to find the effective propagator. By recalling (\ref{gel}) and
using that

\begin{equation}
\int\frac{d^3k}{(2\pi)^3} I(k,p) k_{\alpha\beta}= -\frac{p_{\alpha\beta}}2
\int\frac{d^3k}{(2\pi)^3}I(k,p),
\end{equation}
 we obtain

\begin{eqnarray}
S_3(p)&=&\frac{N}{16\pi} \int d^2\q f(p) [\,- p_{\gamma\b}+2m\,\,
C_{\gamma\beta}]
A^{\b}(p,\q) W^{\gamma}(-p,\theta),
\nonumber\\&&\label{t2p}
\end{eqnarray}

\noindent
where $W^{\gamma}$ is the linear part of the field strength, i.e.,

\begin{equation}
\label{strength}
W^{\gamma}=\frac12 D^{\a}D^{\gamma}A_{\a}=\frac12 D^{\gamma}D^{\a}A_{\a}+ D^2
A^\gamma.
\end{equation}

After some straightforward manipulations Eq. (\ref{t2p}) can be rewritten as

\begin{eqnarray}
\label{t2p1}
S_3(p)&=& \frac{N}{16\pi} \int d^2\q f(p) A^{\b}(p,\q) [ D^2+2m  ]
W_{\b}(-p,\q)\nonumber\\& =& \frac{N}{16\pi} \int d^2\q f(p)[ W^\alpha 
W_\alpha+ 2m 
W^\alpha A_\alpha], 
\end{eqnarray}

\noindent
which is, of course,  invariant under the linearized gauge transformation 
$\delta A^{\a}=D^{\a}K$.
The two terms in the last equality in
Eq. (\ref{t2p1}) are nonlocal versions of the Maxwell and Chern-Simons
actions. In the commutative situation the effective action
also contains nonlocal Maxwell and Chern-Simons terms but in contrast
with the above result in that case the leading small $p$ terms
are local.

For quantization the above gauge freedom must be eliminated. This is
  done by adding to (\ref{t2p}) the following gauge fixing action

\begin{eqnarray}
S_{GF}(p)&=&\frac{N}{32\pi\xi} \int d^2\q f(p)D^{\b}A_{\b}(p,\theta)D^2D^{\a}
A_{\a}(-p,\theta).
\nonumber\\&&
\end{eqnarray}

\noindent
Hence the pure gauge total quadratic action is

\begin{equation}
S_{A^\alpha}(p)=-\frac{N}{32\pi} \int \frac{d^3 p}{(2\pi)^3} d^2 \theta f(p)
A_\alpha (-p,\theta) [  D^\beta D^\alpha (D^2+ 2m) + \frac{1}\xi
D^\alpha D^\beta D^2] A_\beta(p,\theta).
\end{equation}

\noindent
This leads to the following effective propagator
\begin{eqnarray}
\label{progen1}
<A^{\a}(p,\q_1)A^{\b}(-p,\q_2)>&=&\frac{4\pi i}{Nf(p)} \left [
\frac{(D^2-2 m) D^\b D^\a}{p^2(p^2 + 4 m^2)}+\xi\frac{ D^2 D^\a D^\b}{(p^2)^2}
\right ]\delta_{12},
\end{eqnarray}

\noindent
which can also be written as

\begin{eqnarray}
\label{progen}
<A^{\a}(p,\q_1)A^{\b}(-p,\q_2)>&=&\frac{4\pi i}{Nf(p)}\Big [-\frac{
2 m p^{\a\b}}{p^2(p^2+4{ m}^2)}+(\frac{1}{p^2+4{ m}^2}-\frac{\xi}{p^2})
C^{\a\b}\nonumber\\&&
+
\frac{1}{p^2}(\frac{1}{p^2+4{ m}^2}+\frac{\xi}{p^2})p^{\a\b}D^2
+\frac{2 m C^{\alpha\b}}{p^2(p^2+4{ m}^2)}D^2
\Big]
\delta_{12}.
\end{eqnarray}

\noindent
As for low momenta $f(p)\simeq \sqrt{{\tilde p}^2}$ then the effective
propagator (\ref{progen}) behaves as $1/{p^3}$. Nevertheless, as in 
the nonsupersymmetric model, due to
the sine factors in the vertices  no infrared divergence should arise from such
behavior.

Aiming to a detailed investigation of the renormalization properties
of the model we now examine the UV limit of the above propagator.
For high momenta we need to consider only the planar contributions as the 
nonplanar ones decay very rapidly. In this situation 
$f(p)\simeq \pi/{\sqrt{p^2}}$ so that

\begin{equation}
 <A^{\a}(p,\q_1)A^{\b}(-p,\q_2)>\simeq \frac{4i}{N}[\frac{1-\xi}
{({p^2})^{1/2}}
C^{\alpha\beta}+\frac{1+\xi}{(p^2)^{3/2}} p^{\alpha\beta} D^2] \delta_{12}.
\end{equation}

\noindent
For $\xi=-1$ this expression assumes the simpler form

\begin{eqnarray}
\label{prol}
<A^{\a}(p,\q_1)A^{\b}(-p,\q_2)>=\frac{8i}{N}\frac{C^{\a\b}}{(p^2)^{1/2}}
\delta_{12}.
\end{eqnarray}

The action of the  Faddeev-Popov ghosts is
\begin{eqnarray}
S_{FP}=-\frac{N}{32\pi}\int d^3 p d^2\theta f(p)(c'D^2 c-ic'D^\a[ A_\a, c]_*),
\end{eqnarray}

\noindent
yielding the ghost propagator

\begin{eqnarray}
<c'(p,\theta_1) c(-p,\theta_2)>=-i\frac{32\pi}{N}\frac{D^2}{p^2f(p)}\delta_{12}.
\end{eqnarray}

\noindent
A direct consideration of the supergraphs involving ghost loops shows that 
they will begin to contribute only in the $\frac{1}{N^2}$ order.

In the anticommutator case we notice that the two point vertex functions
of $\phi $, $A_\alpha$
and the planar part of the $\eta$ fields
 are the same as we calculated before but the nonplanar part of the
two point vertex function of the
$\eta$ field changes  sign.  Besides that,   the additional effective action

\begin{equation}
S_{A \eta}= -\frac{N}{8\pi} \int_{0}^{1} dx \frac{{\rm e}^{-M\sqrt{{\tilde p}^2}}}{\sqrt{{\tilde p}^2}} {\tilde p}_{\beta\gamma} A^\gamma(-p,\theta) D^\beta
\eta(p,\theta),
\end{equation}

\noindent
coming from the graph in Fig. \ref{Fig6},
is induced, leading to a nonvanishing mixed propagator $<A_\alpha
\eta> $.  From the above expression one sees that any graph containing
the mixed propagator is superficially convergent; thus, to analyze the UV
behavior of the Green functions  we can discard such graphs and use
the same propagators as in commutator case.

\section{Radiative corrections to the two point vertex function of the scalar field}
\label{sec4}
\begin{figure}[ht]
\includegraphics{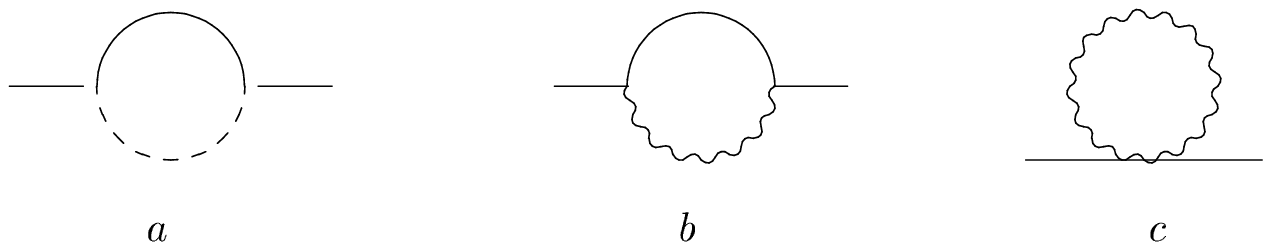}
\caption{Order $1/N$ contributions to the two point vertex function of the
$\varphi$ field. Continuous, dashed and wavy lines represent the
propagators for the $\varphi$, $\lambda$ and $A_\mu$ fields,
respectively.}
\label{Fig5}
\end{figure}

\begin{figure}[ht]
\includegraphics{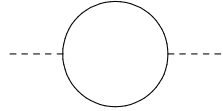}
\caption{Lowest order contribution to the propagator of the auxiliary $\eta$
field. Here the dashed line is for $\eta$ field and the continuous line  for $\phi_a,\bar{\phi}_a$
fields.} 
\label{Fig3}
\end{figure}

\begin{figure}[ht]
\includegraphics{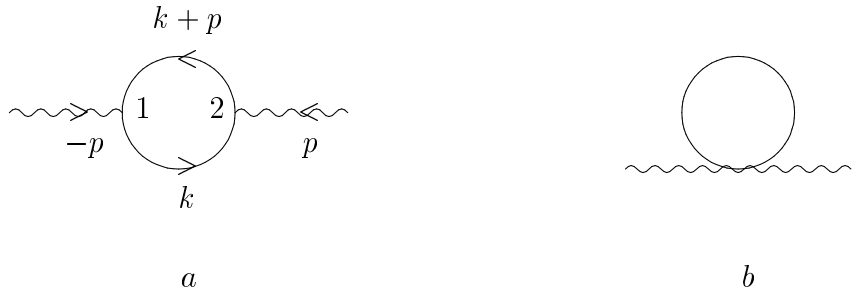}
\caption{Lowest order contributions to the propagator of the auxiliary
$A^\alpha$ field.}
\label{Fig4}
\end{figure}

\begin{figure}[ht]
\includegraphics{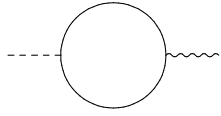}
\caption{A potential  contribution to the two point vertex function of $\eta$ and $A_\alpha$ fields.  
} 
\label{Fig6}
\end{figure}

At the next to leading order in $\frac{1}{N}$ the divergentt 
contributions to the two point vertex function of $\f$ field are given by the graphs
shown in Fig. \ref{Fig5}, where  continuos, wavy and dashed lines now
represent the propagators for the $\phi$, $A_\alpha$ and $\eta$ superfields.  
Using the propagator in Eq. (\ref{peta}) for $\eta$ field, the amplitude for the graph in Fig. \ref{Fig5}$a$ is 

\begin{eqnarray}
i S_{1a}(p)&=&\frac{16\pi}{N}\int\frac{d^3k}{(2\pi)^3}\int d^2\q_1d^2\q_2
\f_a(-p,\q_1)\bar{\f}_a(p,\q_2)
\frac{\sin^2(k\wedge p)}{((k+p)^2+m^2)f(k)(k^2+4m^2)}
\nonumber\\&\times&
(D^2-2m)\delta_{12}(D^2+m)\delta_{12}.
\end{eqnarray}
\noindent
By doing the usual  D-algebra transformations (cf. \cite{sig}) and replacing
$f(k)$ by its asymptotic form $f(k) \approx \pi/\sqrt{k^2}$ we get

\begin{eqnarray}
\label{s31}
i S_{1a}(p)&=&\frac{16}{N}\int\frac{d^3k}{(2\pi)^3}\int d^2\q
\f_a(-p,\q)(D^2-m)\bar{\f}_a(p,\q)
\frac{\sqrt{k^2}\sin^2(k\wedge p)}{((k+p)^2+m^2)(k^2+4m^2)},
\end{eqnarray}

\noindent
which, by power counting is only logarithmically divergent.

The graph shown in  Fig. \ref{Fig5}$b$ contributes
\begin{eqnarray}
\label{s320}
i S_{1b}(p)&=&\frac{4\pi}{N}
\int\frac{d^3k}{(2\pi)^3}\int d^2\q_1 d^2\q_2\frac{\sin^2(k\wedge p)}{f(p-k)}\frac{1}{k^2+m^2}
\Big[\frac{
2 m (p-k)^{\a\b}}{(p-k)^2((p-k)^2+4{ m}^2)}\\&+&
(\frac{1}{(p-k)^2+4{ m}^2}-\frac{\xi}{(p-k)^2})
C^{\a\b}-
\frac{1}{(p-k)^2}\Big(\frac{1}{(p-k)^2+4{ m}^2}
\nonumber\\&+&\frac{\xi}{(p-k)^2}\Big)(p-k)^{\a\b}D^2
+\frac{2 m C^{\a\b}}{(p-k)^2((p-k)^2+4{ m}^2)}D^2\Big]
\delta_{12}\
\nonumber\\&\times&\Big[
(D^2+m)D_{\b 2}\delta_{12}D_{\a}\f_a(p,\q_1)\bar{\f}_a(-p,\q_2)-
D_{\a 1}(D^2+m)\delta_{12}\f_a(p,\q_1)D_{\b}\bar{\f}_a(-p,\q_2)
\nonumber\\&+&
(D^2+m)\delta_{12}D_{\a}\f_a(p,\q_1)D_{\b_2}\bar{\f}_a(-p,\q_2)+
D_{\a 1}(D^2+m)D_{\b 2}\delta_{12}\f_a(p,\q_1)\bar{\f}_a(-p,\q_2)
\Big].\nonumber
\end{eqnarray}

  Superficially
$S_{1b}$ contains linear divergences. However, the UV leading term of
$S_{1b}$, after the D-algebra transformations, turns out to be
proportional to 

\begin{eqnarray} \int\frac{d^3k}{(2\pi)^3}\int d^2\q
\frac{k_{\b\a}\sin^2(k\wedge p)}{(k^2)^{3/2}}
C^{\a\b}\f_a(-p,\q)\bar{\f}_a(p,\q), \end{eqnarray} 

\noindent
which vanishes since
$C^{\a\b}k_{\b\a}=0$.  
Therefore  $iS_{1b}$ in Eq. (\ref{s320}) is only
logarithmically divergent. To obtain this divergent part we 
delete the $4 m^2$ terms in the denominators of (\ref{s320}) and
replace $f(p-k)$ by its asymptotic form. We then have 
the sum of three contributions:

a) The term proportional to $2m$. After the  commutation of $D_{\beta 2}$ with
$D^2$ and the use of $D_{\beta 2}\delta_{12}=-D_{\beta 1}\delta_{12}$ it 
contributes with:
\begin{eqnarray}
\label{s320a}
iS_{1b}^{(1)}&=&\frac{8m}{N}
\int\frac{d^3k}{(2\pi)^3}
\int d^2\q_1 d^2\q_2\sin^2(k\wedge p)\frac{1}{k^2+m^2}
\\&\times&
\Big[
\frac{(p-k)^{\alpha\beta}+C^{\alpha\beta}D^2}{[(p-k)^2]^{3/2}}\Big]
\delta_{12}\nonumber\\&\times&
D_{\alpha 1}D_{\beta 1}(D^2-m)\delta_{12}\f_a(p,\q_1)\bar{\f}_a(-p,\q_2).
\nonumber
\end{eqnarray}
We now apply the identity $D_{\a 1}D_{\b 1}=k_{\a\b}-C_{\a\b}D^2$
which implies in
\begin{eqnarray}
\label{s320b}
iS_{1b}^{(1)}&=&\frac{8m}{N}
\int\frac{d^3k}{(2\pi)^3}
\int d^2\q_1 d^2\q_2\sin^2(k\wedge p)\frac{1}{k^2+m^2}
\\&\times&
\Big[
\frac{(p-k)^{\alpha\beta}+C^{\alpha\beta}D^2}{[(p-k)^2]^{3/2}}\Big]
\delta_{12}\nonumber\\&\times&
[k_{\alpha\beta}D^2+C_{\alpha\beta}k^2-
mk_{\alpha\beta}+mC_{\alpha\beta}D^2]
\delta_{12}\f_a(p,\q_1)\bar{\f}_a(-p,\q_2).
\nonumber
\end{eqnarray}
After contracting the loop into a point we arrive at the following
divergent correction
\begin{eqnarray}
\label{s320c}
iS_{1b}^{(1)}&=&\frac{8m}{N}
\int\frac{d^3k}{(2\pi)^3}
\int d^2\q_1 d^2\q_2\sin^2(k\wedge p)\frac{1}{k^2+m^2}
\\&\times&
\Big[
\frac{-k^{\alpha\beta}k_{\alpha\beta}+
C^{\alpha\beta}C_{\alpha\beta}k^2}{[(p-k)^2]^{3/2}}\Big]
\delta_{12}
\f_a(p,\q_1)\bar{\f}_a(-p,\q_2).
\nonumber
\end{eqnarray}
Since $-k^{\alpha\beta}k_{\alpha\beta}+
C^{\alpha\beta}C_{\alpha\beta}k^2=2k^2-2k^2=0$ the term 
proportional to $2m$ gives zero contribution.

b) The term proportional to $(\xi+1)$ contributes with
\bea
iS_{1b}^{(2)}(p)&=&-\frac{8}{N}\f_a(-p,\q)(3D^2-m)\bar{\f}_a(p,\q)
(1+\xi)\int\frac{d^3k}{(2\pi)^3}
\frac{\sin^2(k\wedge p)}{(k^2+m^2)((p-k)^2)^{1/2}}.
\eea

c) The term proportional to $(\xi-1)$ contributes with
\bea
iS_{1b}^{(3)}(p)&=&-\frac{8}{N}\f_a(-p,\q)(D^2+m)\bar{\f}_a(p,\q)
(1-\xi)\int\frac{d^3k}{(2\pi)^3}
\frac{\sin^2(k\wedge p)}{(k^2+m^2)((p-k)^2)^{1/2}}.
\eea

By adding the UV leading (logarithmically divergent) parts of 
  $iS_{1b}^{(2)}$,  $iS_{1b}^{(3)}$
the total divergent contribution to  $iS_{1b}$ is equal
to 

\begin{eqnarray}
\label{s32}
iS_{1b}(p)&=&-\frac{16}{N}\int \frac{d^3p}{(2\pi)^3}\int d^2\q
[(2+\xi)\f_a(-p,\q)D^2\bar{\f}_a(p,\q)-m\xi
\f_a(-p,\q)\bar{\f}_a(p,\q)]
\nonumber\\&\times&
\int\frac{d^3k}{(2\pi)^3}
\frac{\sin^2(k\wedge p)}{((k+p)^2+m^2)\sqrt{k^2}}.
\end{eqnarray}

The linearly divergent part of the graph given in
 Fig. \ref{Fig5}$c$ in { any} gauge, after the D-algebra
 transformations is

\begin{eqnarray}
&\propto&\int\frac{d^3k}{(2\pi)^3}\int d^2\q
\frac{k_{\b\a}\sin^2(k\wedge p)}{(-k^2)^{3/2}}
C^{\a\b}\f_a(-p,\q)\bar{\f}_a(p,\q),
\end{eqnarray}

\noindent
which vanishes being  proportional to $C^{\a\b}k_{\b\a}=0$. 
However, there are logarithmically divergent contributions  given by
\begin{eqnarray}
\label{s33}
iS_{1c}(p)&=&\frac{32}{N}m\int \frac{d^3p}{(2\pi)^3}\int d^2\q
\f_a(-p,\q)\bar{\f}_a(p,\q)
\nonumber\\&\times&
\int\frac{d^3k}{(2\pi)^3}
\frac{\sin^2(k\wedge p)}{((k+p)^2+m^2)\sqrt{k^2}},
\end{eqnarray}

\noindent
coming from the graph in Fig. \ref{Fig5}$c$.

We conclude that
the contribution to the effective action arising from the 
sum of (\ref{s31}), (\ref{s32}) and 
(\ref{s33}) is also free of dangerous UV/IR mixing
and has the form

\begin{eqnarray}
iS_{\f\bar{\f}}(p)&=&-\frac{16}{N}(1+\xi)
\int \frac{d^3p}{(2\pi)^3}\int d^2\q
\f_a(-p,\q)(D^2-m)\bar{\f}_a(p,\q)
\nonumber\\&\times&
\int\frac{d^3k}{(2\pi)^3}
\frac{\sin^2(k\wedge p)}{((k+p)^2+m^2)\sqrt{k^2}}+fin,
\end{eqnarray}

\noindent
where $fin$ denotes the remaining terms which are UV
finite and possess at most a logarithmic UV/IR infrared divergence
(actually, because of the sine factor no infrared divergence appears). 
We see that the quadratic UV/IR infrared divergence that occurred in the
nonsupersymmetric version of the model, discussed in Section \ref{fermion}, has disappeared under the present
supersymmetrization.
After integration of the planar part,  $S_{\f\bar{\f}}$
becomes

\begin{eqnarray}
\label{squa}
S_{\f\bar{\f}}(p)=-\frac{4(1+\xi)}{N\pi^2\epsilon}\int\frac{d^3p}{(2\pi)^3}\int d^2\q
\f_a(-p,\q)(D^2-m)\bar{\f}_a(p,\q)
+fin.
\end{eqnarray}

\noindent
This divergence can be canceled with the help of an appropriate
counterterm which implies in the following wave function renormalization constant for
kinetic term $\f_a (D^2-m)\bar{\f}_a$:

\begin{eqnarray}
\label{ren}
Z=1+\frac{4(1+\xi)}{\pi^2N\epsilon},
\end{eqnarray}

\noindent
so that in the gauge $\xi=-1$ the correction is finite.

\section{The general structure of divergences and the absence of dangerous UV/IR
mixing}
\label{sec5}
We have explicitly verified, that the two point vertex functions of
the $\phi$ field up to first order in $\frac{1}{N}$ do not produce
nonintegrable divergences. To further clarify the issue of
renormalizability up to order $1/N$, we start by calculating the
superficial degree of divergence $\omega$ of an arbitrary graph
$\gamma$. To that end, let us denote the number of vertices
$iA^{\a}*(D_{\a}\f_a*\bar{\f}_a-D_{\a}\bar{\f}_a*\f_a)$ by $V_1$, of
$A^{\a}*A_{\a}*\f_a*\bar{\f}_a$ by $V_2$,  of
$\eta*\bar{\f}_a*\f_a$ by $V_3$, and of $f(p)c'\ast D^\alpha[A_\alpha, \, c]_
\ast$ by $V_c$.  Furthermore let
$P_{\f},P_{A},P_{\eta},P_c$ be the number of propagators
$<\f_a\bar{\f}_b>,<A^{\a}A^{\b}>\sim\frac{D^2}{k^2}$ and $<\eta\eta>$,
$<cc'>\sim\frac{D^2}{\sqrt{k^2}}$, respectively.  Each loop
contributes to $\omega$ with 2 (3 for the integral over $d^3k$, $-1$
because the contraction of a loop into a point decreases the number of
$D^2$-factors which could be converted to momenta by 1). Each $\f_a$
or $A_{\a}$  propagator contributes with $-1$ while each vertex
$V_1$ brings $\hf$ since it contains one spinor derivative and each vertex $V_c$ brings $-\hf$ due to the factor $f(p)$ . Therefore, $\omega$
 is

\bea
\label{o1}
\o=2L-P_{\f}-P_A+\hf V_1 -\hf V_c.
\eea

\noindent
where $L$ designates the number of loops in $\gamma$.
By using the well known topological identity $L+V-P=1$,  this becomes
\bea
\label{o2}
\o=2+ P_A+P_{\f}+2(P_{\eta}+P_c)-\frac{3}{2} V_1- \frac{5}{2} V_c-2(V_2+V_3)
-P_A-P_{\f}.
\eea

The number of propagators may be expressed in terms of the number of 
the external lines $E_{\f},\,E_A,\, E_{\eta},\, E_c$ and of the
 total number of fields $N_{\f},\, N_A,\, N_{\eta},\, N_c$ used to construct $\gamma$ 
 as

\bea
\label{o3}
P_{\f}=\hf(N_{\f}-E_{\f});\quad P_A=\hf(N_A-E_A);\quad P_{\eta}=\hf(V_3-E_{\eta});\quad
P_c=\hf(N_c-E_c) .
\eea

\noindent
It is then  easy to verify that

\bea
\label{o4}
N_{\f}=2(V_1+V_2+V_3);\quad N_A=V_1+2V_2+ V_c;\quad N_{\eta}=V_3; \quad N_c= 2 V_c.
\eea

By replacing Eqs. (\ref{o4}) and (\ref{o3}) into (\ref{o2}),
and after taking into account that $\o$  decreases  by
$\frac{N_D}2$ when $N_D$ supercovariant derivatives are
moved to the external lines, one arrives at

\bea \o=2-\hf(E_{\f}+E_A)-E_{\eta}-E_c-\hf N_D. 
\eea 

\noindent
We immediately see
that $\o$ in the theory cannot be larger than one (it would be two
only for vacuum supergraphs but these contributions vanish due to the
properties of the integral over Grassmann coordinates \cite{SGRS}).
We  also note that $E_{\f}$ must be even in order to have an (iso)scalar
contribution. By the same reason,  $E_A$  must either be  even or if
not it must be accompanied by an odd number of spinor
supercovariant derivatives.

The case $\o=1$ corresponds to $E_{\f}=2$, or $E_A=2$, or
$E_{\eta}=1$, or $E_A=N_D=1$
(with the number of all other external lines in each case being
zero). However, we have already proved that the graphs with $E_{\f}=2$  (depicted in
Fig. \ref{Fig5}) 
are at most  logarithmically divergent, that the sum of the graphs with
$E_A=2$ (which are depicted in Fig. \ref{Fig4}) is finite and 
contributes only to the effective
propagator of the gauge field. Besides that, the
graph with $E_{\eta}=1$ is a tadpole graph which automatically
vanishes in the commutator case, whereas in the anticommutator case
its effect is only to fix $m$ as being the mass of the $\phi$ superfield
(compare with \cite{sig}). As for
the graph corresponding to $E_A=N_D=1$, which is formally linearly
divergent, its contribution is proportional to $\int d^5 z
D^{\a}A_{\a}$ which is of course zero.

From this discussion, we see that, up to the leading order of the
$\frac{1}{N}$ expansion, all divergences in the theory are only
logarithmic.  It means that the quantum corrections in the theory are, up to
this order,  free from nointegrable  infrared UV/IR  singularities.  We  
hope that a similar situation takes place at higher orders in the
$\frac{1}{N}$ expansion.

There are more possibilities if $\o=0$, namely $E_A=4$, or $E_{\f}=4$, or
$E_A=E_{\f}=2$, or $E_{\eta}=1,E_{\f}=2$, 
or $E_{\eta}=1,E_A=2$, or 
$E_{\eta}=2$, or $E_A=1,E_{\f}=2,E_D=1$, or $E_A=3,E_D=1$. The cases
with either $E_\f=4$ or $E_A=4$ or $E_A=3$ or $E_\eta=1$, $E_A=2$ are
particularly dangerous because they are
 potentially logarithmically divergent but there is no available counterterm to absorb these divergences. However, one can explicitly verify that in all
these cases the integrands associated to the divergent parts are odd in the loop momentum and therefore vanish under symmetric integration.
Thus up to leading order of the  $\frac1N$ expansion
only the cases $E_{\eta}=1,E_{\f}=2$, 
or $E_{\phi}=2,E_A=2$, or  $E_A=1,E_{\f}=2,E_D=1$ imply in divergences.

This means that we can construct effective interaction terms for
an effective Lagrangian of the gauge field $A^{\a}$ which are finite
and proportional to $\int d^5 z \frac{1}{\sqrt{\Box}}(DA)^2 A^2$ and
$\int d^5 z (DA) A^2$, which are needed to complete the induced noncommutative
Maxwell and Chern-Simons Lagrangians.  The graph with two external
$\eta$  fields of  order $N$ is given by Fig. \ref{Fig3}, and we already showed that it
is finite. As for the subleading graphs with two external $\eta$
fields they could only modify the effective propagator of $\eta$ field
in higher orders of the $\frac{1}{N}$ expansion.

In the commutator case, due to the invariance of the action (\ref{acti}) under charge
conjugation, the contributions proportional to $\eta A^{\a}A_{\a}$
vanish in any finite order of the expansion. In particular, at the
first order in $\frac{1}{N}$  this result can be seen
directly as it turns out to be proportional to $\int d^2\q d^3p_1 d^3
p_2 A^{\a}(p_1,\q)A_{\a}(p_2,\q)
\eta(-p_1-p_2,\q)\sin(p_1\wedge p_2)$, which evidently vanishes.

To sum up, in the leading order of $1/N$ the only logarithmic divergences in the theory are those
ones proportional to $\f_a\bar{\f}_a$, which give origin to the wave
function renormalization of the $\phi$ field, those ones proportional
to $\eta\f_a\bar{\f}_a$, which, by a method similar to that  employed
in \cite{sig}, can be shown to have the same Moyal structure as the
corresponding vertex in the classical action (both in the 
commutator and anticommutator cases),
and  those which are  proportional to
$\f_a\bar{\f}_aA^{\a}A_{\a}$ and to $A^{\a}(D_{\a}\f_a) \bar{\f}_a$,
$A^{\a}\f_a (D_{\a}\bar{\f}_a)$.

It is easy to verify that the Moyal structure of the quantum
corrections proportional to $A^{\a}(D_{\a}\f_a) \bar{\f}_a$,
$A^{\a}\f_a (D_{\a}\bar{\f}_a)$ is preserved. For example, 
in the commutator case each supergraph
in such quantum corrections contains an odd number of the triple
vertices, and therefore they will furnish a product of an odd number
of sine factors; thus, as  in \cite{sig}, we
find that the planar contribution could have only the form
$\sin(p_1\wedge p_2)$ where $p_1$ and $p_2$ are two of the three
incoming momenta. Such  phase factor also reproduces the corresponding Moyal structure in the classical action.
However, an analogous proof
of the same fact for the quartic correction
$\f_a\bar{\f}_aA^{\a}A_{\a}$ is much more complicated. This problem
will be considered elsewhere.

\section{Summary}
\label{sec6}
In this paper we studied the minimal and supersymmetric inclusion of
fermions in the pure noncommutative $CP^{N-1}$ model. Although for
both situations a great improvement in the long distance behavior of the gauge
two point vertex function was achieved, the case of minimal coupling still
presented an infrared nonintegrable singularity in the self-energy of
the basic scalar field. To evade this problem the supersymmetric
extension was also considered and we proved that, at least to  $1/N$ 
order, the supersymmetric model is free from a dangerous UV/IR
mixing. This is a strong indication that this supersymmetric model has a
consistent perturbative expansion.  The theory
exhibits very nontrivial properties as the generation of  a dynamics
for the spinor connection superfield, with both the Maxwell
and the Chern-Simons terms being generated by the quantum
contributions. The ghost superfields which are generated also possess 
nontrivial dynamics; however, they contribute to the quantum
corrections only in $\frac{1}{N^2}$ and higher orders.

The analysis of the ultraviolet behavior  unveiled some
interesting aspects of the renormalization program for the two
versions of the model. In both cases considered, the model 
turns out to be renormalizable since the use   of a commutator
or an anticommutator does not change the planar part of the
amplitudes. All ultraviolet divergences are logarithmic and 
can be eliminated by adequate counterterms (the Moyal
structure of  the $\bar \phi_a \phi_a A^\alpha A_\alpha$ vertex still needs
further analysis).
Similarly to the noncommutative nonlinear sigma model, nontrivial wave function
renormalizations for the auxiliary $\eta$ and $A_\alpha$ fields are expected
\cite{sig}.

The wave
function renormalization constant for the scalar superfield was shown
to be gauge dependent 
whereas, due to charge conjugation invariance,
the mixed two point vertex function of the $A_\alpha$ and $\eta$ fields  vanishes in the commutator
case.

  A further development of the model
could consist in a more detailed investigation of the higher orders of
the $\frac{1}{N}$ expansion. Also, it could be interesting to develop
the extended supersymmetric generalization of this model by analogy
with the $\N=2$ super-Yang-Mills theory containing gauge and matter
multiplets in $\N=1$ description.

\section{Acknowledgments} 

This work was partially supported by Funda\c{c}\~{a}o de Amparo 
\`{a} Pesquisa do Estado de S\~{a}o Paulo (FAPESP) and Conselho 
Nacional de Desenvolvimento Cient\'{\i}fico e Tecnol\'{o}gico (CNPq). 
H. O. G. also acknowledges support from PRONEX
under contract CNPq 66.2002/1998-99. A. Yu. P. has been supported by
FAPESP, project No. 00/12671-7.

\newpage

\end{document}